\documentstyle[aps,prb]{revtex}
\draft

\begin{document}

\title{Quantum Monte Carlo in the Interaction Representation ---
Application to a Spin-Peierls Model}

\author{A. W. Sandvik}
\address{Department of Physics, University of Illinois at Urbana-Champaign,
1110 West Green Street, Urbana, Illinois 61801}
\author{R. R. P. Singh}
\address{Department of Physics, University of California, Davis, 
California 95616}
\author{D. K. Campbell}
\address{Department of Physics, University of Illinois at Urbana-Champaign,
1110 West Green Street, Urbana, Illinois 61801}

\date{June 5, 1997}
\maketitle{}

\begin{abstract}
\end{abstract}

\section{Introduction}

Over the last two decades, discrete Euclidean path integrals constructed 
using the so called Trotter decomposition\cite{trotter,suzuki1} have been 
widely used as the starting point for quantum Monte Carlo (QMC) simulations of
lattice models at finite temperature.
\cite{reviews,suzuki2,spinqmc,worldline} 
In these ``worldline'' methods,\cite{worldline} the discretization 
$\Delta\tau$ in imaginary time introduces a systematic error in computed 
quantities, which in principle can be eliminated by carrying out simulations 
for different discretizations and extrapolating 
to $\Delta\tau =0$. The error typically \cite{suzuki1,fye} scales as 
$(\Delta\tau)^2$. However, algorithms can also be constructed which 
are associated with no inherent systematic errors, thus eliminating the need 
for multiple simulations and extrapolations. Such a QMC scheme, applicable to 
the ferromagnetic spin-1/2 Heisenberg model, was 
first devised by Handscomb as early as in 1961.\cite{handscomb} This method
is based on a series expansion of the density matrix operator 
exp$(-\beta\hat H)$, which in
the case of the Heisenberg model can be written in terms of products of 
permutation operators. Their traces can be evaluated exactly and are positive
definite. One can then carry out importance sampling in the space of operator 
sequences, and obtain results exact to within statistical errors. Handscomb's 
method is not directly applicable to other models,\cite{exchangemodel} 
not even for the antiferromagnetic Heisenberg model for
which the scheme breaks down due to the non-positive-definiteness of the 
traces.\cite{lyklema} There was therefore not much follow-up on Handscomb's 
pioneering efforts, and there was little progress towards practical QMC 
algorithms until Suzuki proposed the use of the Trotter formula in this 
context.\cite{suzuki1} Several methods based on this controlled approximation 
were subsequently developed.\cite{spinqmc,worldline,detqmc}
Variants of Handscomb's method were also later developed, for 
the antiferromagnetic Heisenberg model by Lee {\it et al.},\cite{lee} and 
for the XY-model by Chakravarty and Stein,\cite{chakravarty} but this 
type of simulation scheme was for a long time still perceived as 
fundamentally limited by its reliance on the special properties of spin-1/2 
operators. \cite{suzuki3,negele} However, the generalizations of Handscomb's 
method developed for $S > 1/2$ spin models by Sandvik and Kurkij\"arvi,
\cite{sse1} and for 1D Hubbard-type models by Sandvik,\cite{sse2} have now
clearly demonstrated that non-approximate algorithms based on ``stochastic 
series expansion'' (SSE) can in principle be constructed for any lattice 
model. In this scheme, the basic limitation of the earlier formulations of 
Handscomb's method is overcome by expanding the traces as diagonal matrix 
elements in a suitable chosen basis. The importance sampling is then carried 
out in a space of basis states and operator sequences.\cite{sse1,sse2} In 
practice, this type of method is of course still limited to models for which 
a positive definite weight function can be achieved. The cases for which this 
is possible coincide with those for which the weight is positive definite also
in the standard Trotter-based path integral formulations. In fact, despite 
the different starting points of the two approaches, the SSE configuration 
space is strongly related to an Euclidean path integral. Many characteristics
are therefore shared, including the range of practical applicability, the 
types of observables accessible for evaluation, and the scaling of the 
computation time with the system size and the inverse temperature $\beta$. 
The main advantage of SSE is of course the absence of systematic errors. 
It should be noted that in order to eliminate the Trotter error in worldline 
calculations, simulations have to be carried out for several sufficiently
small discretizations $\Delta\tau$. Since the computation time scales as 
$1/(\Delta\tau)^3$,\cite{prokofev} the SSE approach can in practice be 
considerably more efficient than the worldline method in cases where 
completely unbiased results are needed.

Recently, other approaches to exact QMC algorithms have been proposed. Beard
and Wiese succeeded in formulating a worldline algorithm for the spin-1/2 
Heisenberg model directly in the $\Delta\tau \to 0$ limit.\cite{beard} 
Constructed within the framework of a non-Metropolis sampling scheme with 
global ``loop-cluster'' updates previously developed by Evertz {\it et al.}
\cite{evertz} (a generalization of the classical cluster spin algorithm 
\cite{swendsen}), this method also has the added advantage of significantly
shorter autocorrelation times. Concurrently, Prokof`ev, Svistunov and
Tupitsyn suggested the use of the standard perturbation expansion as a basis 
for a QMC path integral. \cite{prokofev} For a finite system at finite 
temperature the series converges to an exact result for a finite number of 
terms. A scheme involving a novel class of local updates was suggested for 
efficiently sampling the continuous-time paths.

It is clear that the methods by Beard and Wiese\cite{beard} and Prokof`ev 
{\it et al.}\cite{prokofev} are strongly related to each other, involving
the same configuration space of world lines in continuous imaginary time,
but differing in the sampling procedures. Continuous-time path integrals also 
have many properties in common with the SSE sum.\cite{sse1,sse2} Notably, a 
transition event in imaginary time corresponds directly to the presence of an 
off-diagonal operator in the SSE operator string. Here we discuss this 
connection in detail, and introduce a simple modification of the SSE algorithm
for simulations in the interaction representation. This formulation can be 
expected to be more efficient than standard SSE in cases where the diagonal 
part of the Hamiltonian dominates. In order to explore the properties of the 
new method, we study the spin-1/2 Heisenberg chain, as well as a spin chain 
including couplings to dynamic (fully quantum mechanical) phonons. We consider
a coupling via a linear modulation of the spin exchange by a local 
dispersionless oscillator (Einstein phonon). The system undergoes a 
spin-Peierls (dimerization) transition at zero temperature. A detailed study 
of the model, and its relevance for understanding the magnetic properties of 
the recently discovered spin-Peierls compounds GeCuO$_{\rm 3}$ 
(Ref.~\onlinecite{gediscovery}) and $\alpha'-$NaV$_{\rm 2}$O$_{\rm 5}$
(Ref.~\onlinecite{nadiscovery}), will be presented elsewhere.\cite{awsdkc}
Here we only consider a single set of model parameters, in the general regime 
expected to be of physical relevance, and illustrate the use of the new method
to calculate a variety of physical observables, both at finite temperature 
and in the limit $T \to 0$. Based on the results, we conclude that the effects
of dynamic phonons cannot be neglected in quantitative descriptions of 
materials such as those mentioned above.

The outline of the rest of the paper is the following: In Sec.~II we review 
the formalism of the SSE method. The perturbation expansion in the 
interaction representation and its relation to the SSE series are discussed 
in Sec.~III. In Sec.~IV we implement an interaction representation algorithm 
for the Heisenberg chain, and discuss the performance of the method. In 
Sec.~V we consider the spin-phonon model. Readers interested mainly in the 
new results for this model are advised to skip directly to the introductory 
part of Sec.~V, and then go directly to the results section V-B. The discussion
there does not rely heavily on the previous, more technical parts of the paper.
Sec.~VI concludes with a summary and outlook for further developments and 
applications of the new QMC algorithm, in particular to various models 
including phonons.

\section{Stochastic Series Expansion}

Here we review the general formalism of the SSE method, needed as a basis
for the discussion in the following sections. More details of the 
algorithm have been described elsewhere.\cite{book} Some recent applications 
to spin systems and 1D fermion systems are listed in 
Refs.~\onlinecite{ssespins,nmr,sss,ssefermions,cuo}.

The starting point for evaluating an operator expectation value at inverse
temperature $\beta$,
\begin{equation}
\langle \hat A \rangle = {1\over Z}
{\rm Tr}\lbrace \hat A {\rm e}^{-\beta \hat H} \rbrace ,\quad
Z = {\rm Tr}\lbrace {\rm e}^{-\beta \hat H} \rbrace ,
\label{za}
\end{equation}
is to Taylor expand exp$(-\beta \hat H)$ and write the traces as sums 
over diagonal matrix elements in a basis $\lbrace |\alpha \rangle\rbrace$.
The partition function is then
\begin{equation}
Z = \sum\limits_{\alpha}\sum\limits_{n=0}^\infty {(-\beta)^n\over n!}
\langle \alpha | \hat H^n | \alpha \rangle .
\end{equation}
The Hamiltonian is next written as
\begin{equation}
\hat H= \sum\limits_{b=1}^M \hat H_b,
\label{hsum}
\end{equation}
where the operators $\hat H_b$ have the ``non-branching'' property,
\begin{equation}
H_b |\alpha \rangle = h_b(\alpha,\beta)|\beta \rangle ,
\label{requirement}
\end{equation}
where $|\alpha \rangle$ and $|\beta \rangle$ are both basis states in the 
chosen representation. Each power $\hat H^n$ is now expanded as a sum over all 
possible products of $n$ of the operators $\hat H_b$. With $S_n$ denoting 
an index sequence referring to the operators in the product (the operator 
string),
\begin{equation}
S_n = (b_1 ,\ldots ,b_n), \quad b_i \in \lbrace 1,\ldots ,M \rbrace ,
\end{equation}
the partition function becomes
\begin{equation}
Z = \sum\limits_{\alpha} \sum\limits_{n=0}^\infty \sum\limits_{S_n}
{(-\beta)^n \over n!}
\Bigl \langle \alpha \Bigl | \prod\limits_{i=1}^n \hat H_{b_i}
\Bigr | \alpha \Bigr \rangle .
\label{ssepartition}
\end{equation}

Eqs.~(\ref{hsum}) and (\ref{requirement}) of course represent a completely 
general formal device. In practice, one typically chooses the basis so as to 
make the sum (\ref{hsum}) as simple as possible. For example, for the 
Heisenberg model,
\begin{equation}
\hat H = J\sum\limits_{\langle i,j \rangle} {\bf S}_i \cdot {\bf S}_j,
\end{equation}
the eigenstates of $S^z_i$ can be chosen; 
$|\alpha\rangle = |S^z_1,\ldots,S^z_N\rangle$. Writing the Hamiltonian as
\begin{equation}
\hat H = J\sum\limits_{\langle i,j \rangle} 
\bigl [ S^z_iS^z_j + \hbox{$\rm 1\over 2$}(S^+_iS^-_j + S^-_iS^+_j ) \bigr ],
\end{equation}
all the two-spin operators
satisfy the requirement (\ref{requirement}). For $S=1/2$,
$S^+_iS^-_j + S^-_iS^+_j$ satisfy (\ref{requirement}) and can be 
considered as a single operator, whereas for $S > 1/2$ the two terms have to 
be treated as separate operators. For tight-binding fermion or boson 
models, the real-space occupation number basis is typically chosen. For 
fermions and hard-core bosons the hopping operator $c^+_ic_j + c^+_jc_i$ 
satisfies (\ref{requirement}), whereas for unconstrained bosons the terms
again qualify only individually.

For a finite system at finite $\beta$, the lengths $n$ of the operator strings 
contributing significantly to the partition function are restricted to 
a finite range. In a Monte Carlo simulation of the series expansion, terms 
$(\alpha,S_n)$ are sampled with a probability proportional to the weight 
function corresponding to Eq.~(\ref{ssepartition}):
\begin{equation}
W(\alpha,S_n)={(-\beta)^n \over n!}
\Bigl \langle \alpha \Bigl | \prod\limits_{i=1}^n \hat H_{b_i}
\Bigr | \alpha \Bigr \rangle .
\label{wn}
\end{equation}
Here it will be assumed that $W(\alpha,S_n)$ is positive definite, which
of course is not always the case. With a non-positive-definite weight, 
simulations can in principle still be carried out using $|W|$,\cite{worldline} 
but in practice the statistical fluctuations of calculated expectation values
diverge if the positive and negative contributions almost cancel each other, 
which they do exponentially both with increasing system size and decreasing
temperature (the infamous sign problem).\cite{sign1,sign2} As already 
discussed in the Introduction, this is the most severe limitation of the 
method --- shared also by standard techniques such as the worldline method
(in the case of fermions, ``determinant methods''\cite{detqmc} typically 
are more effective in dealing with the sign problem \cite{sign1}). Still, the 
class of models for which a positive definite $W$ can be achieved is 
significant enough to motivate the continuing development of more efficient 
QMC methods for their study.

Proceeding as in the derivation of (\ref{ssepartition}), the numerator
in (\ref{za}) corresponding to a given operator $\hat A$ of interest is
also expanded. If the expectation value can then be cast into the form
\begin{equation}
\langle \hat A \rangle =
{\sum_{\alpha} \sum_n \sum_{S_n} A(\alpha,S_n)W(\alpha,S_n) \over
\sum_{\alpha} \sum_n \sum_{S_n}W(\alpha,S_n) },
\end{equation}
the simulation estimate of $\langle \hat A \rangle$ is given by the 
average of the estimator $A(\alpha,S_n)$ over the sampled configurations:
\begin{equation}
\langle \hat A \rangle = \langle A(\alpha,S_n) \rangle .
\end{equation}
The formally simplest observable within the framework of the SSE method
is the internal energy, $E = \langle \hat H\rangle$. As in Handscomb's 
original formulation, the estimator involves only the power $n$; 
\cite{handscomb,sse1}
\begin{equation}
E = - \langle n\rangle/\beta .
\label{energy}
\end{equation}
This, in combination with the expression for the heat capacity,
\cite{handscomb,sse1}
\begin{equation}
C = \langle n^2\rangle - \langle n\rangle^2 - \langle n\rangle ,
\end{equation}
shows that the terms contributing significantly are of length $\sim \beta N$
(at low temperatures), where $N$ is the system size. A derivation of 
(\ref{energy}) will be discussed in the next section.

Already at the level of Eq.~(\ref{ssepartition}), the close relationship 
between SSE and a standard Euclidean path integral is evident. The 
operator string defines a set of propagated states $|\alpha (p) \rangle$, 
$p=0,\ldots,n$:
\begin{equation}
|\alpha (p) \rangle \sim \prod\limits_{i=1}^p \hat H_{b_i}
|\alpha \rangle ,\quad 
|\alpha (0) \rangle  = |\alpha \rangle .
\label{alphap}
\end{equation}
A nonzero weight (\ref{wn}) implies the periodicity condition
$|\alpha (0) \rangle = |\alpha (n) \rangle$. The propagation index $p$ plays
a role analogous to imaginary time in a standard path integral. The exact
relation to imaginary time can be obtained by deriving an expression for a 
time-dependent correlation function.\cite{sse2} Consider two diagonal operators
$\hat A_i$ and $\hat A_j$. In a given configuration $(\alpha,S_n)$, their 
eigenvalues in the states $| \alpha (p) \rangle$ are denoted $a_i[p]$ and 
$a_j[p]$. One can show that the time-dependent correlation function 
$C_{ij}(\tau) = \langle \hat A_j(\tau)\hat A_i(0) \rangle$, where
${\rm e}^{\tau H}\hat A_j{\rm e}^{-\tau H}$, is given by \cite{sse2}
\begin{equation}
C_{ij}(\tau) =
\biggl\langle \sum\limits_{m=0}^n {n \choose m}
{\tau^m (\beta -\tau )^{n-m} \over \beta ^n } \bar C_{ij}(m) \biggr\rangle .
\label{taudia}
\end{equation}
Here $\bar C_{ij}(m)$ is a correlator between states separated by $m$
propagations:
\begin{equation}
\bar C_{ij}(m) =
{1\over n+1} \sum\limits_{p=0}^{n} a_j[p+m]a_i[p].
\end{equation}
The periodicity of the propagated states of course implies that $a_j[p+n]=
a_j[p]$. In the equal-time case, only $m=0$ contributes to (\ref{taudia}), 
and $C_{ij}(0)$ is simply given by $a_i[p]a_j[p]$ averaged over $p$:
\begin{equation}
C_{ij}(0) = \Bigl \langle
{1\over n+1}\sum\limits_{p=0}^n a_j[p]a_i[p] \Bigr \rangle .
\label{diacorr}
\end{equation}
Eq.~(\ref{taudia}) shows that an imaginary time separation $\tau$ 
corresponds to a distribution of separations $m$ between propagated states, 
peaked around $m=n\tau/\beta$. Hence, the propagation index $p$ in the SSE 
method indeed is closely related to the time in an Euclidean path integral.

As already discussed above, the range of contributing powers $n$ is limited
in practice. One can therefore explicitly truncate the series expansion
at some maximum power $n=L$, large enough to introduce only an exponentially
small, completely negligible error. By inserting $L-n$ unit operators
in the operator strings, a configuration space is obtained for which 
the sequence length formally is {\it fixed} at $L$. Defining the unit
operator $\hat H_0 = I$, the summation over $n$ in (\ref{ssepartition}) is 
implicitly included in the summation over all sequences $S_L$, if the
range of allowed indices is extended to include also $b_i=0$. The weight
(\ref{wn}) has to be divided by ${L \choose n}$, in order to compensate for 
the number of different ways of inserting the unit operators, resulting in 
\cite{sse1}
\begin{equation}
W(\alpha ,S_L),
{(-\beta)^n (L-n)!\over L!}
\Bigl \langle \alpha \Bigl | \prod\limits_{i=1}^L \hat H_{b_i}
\Bigr | \alpha \Bigr \rangle ,
\label{wl}
\end{equation}
where $n$ now denotes the number of non-$0$ indices in $S_L$. This 
fixed-length formulation is useful for the construction of an efficient 
sampling scheme for the sequences. For purposes of measuring operator 
expectation values, one can still use the expressions discussed above, with 
the sequences $S_n$ obtained by omitting all the zeros in the generated $S_L$.

In order to ensure a sufficiently high truncation $L$, the power $n$ is 
monitored during the equilibration part of the simulation. If $n$ exceeds some
threshold value $L-\Delta_L$, the sequence is augmented with, e.g., 
$2\Delta_L$ randomly positioned unit operators, corresponding to 
$L \to L + 2\Delta_L$. With $\Delta_L \approx L/10$, this procedure typically 
converges rapidly to a proper $L$. During a subsequent simulation (of 
practical duration), $n$ never reaches $L$. The truncation is therefore no 
approximation in practice.

The details of the Monte Carlo sampling procedures of course depend on
the model under consideration. Here only some general principles will be
discussed. The operators $\hat H_{a}$ can be divided into two classes; 
diagonal and off-diagonal. There are no {\it a priori} constraints on the 
number of diagonal operators that can appear in $S_L$. The probability 
of a diagonal operator $\hat H_{\rm dia}$ at a position $p$ is only 
determined by the state $|\alpha (p-1) \rangle$ on which it operates. The 
general strategy for inserting and removing diagonal operators is to attempt 
substitutions with the unit operator $\hat H_{0}$ introduced in the 
fixed-length scheme (note again that $\hat H_{0}$ is not part of the 
Hamiltonian):
\begin{equation}
\hat H_{0} \leftrightarrow \hat H_{\rm dia}.
\label{single}
\end{equation}
This can be attempted consecutively at all positions in $S_L$. The weight 
change needed for calculating the Metropolis or heat-bath acceptance 
probability involves only the matrix element $\langle \alpha (p-1) 
|\hat H_{\rm dia}| \alpha (p-1) \rangle$ 
and the prefactor $(-\beta)^n (L-n)!$, with $n$ changing by $\pm 1$. With
$|\alpha (0) \rangle$ stored initially, the subsequent states can be generated
one-by-one as needed during the updating process. 

Suitable constants have to be added to the diagonal operators in order to make
all the eigenvalues of $-\beta \hat H_{\rm dia}$ positive. According to
Eq.~(\ref{wl}), the presence or absence of a sign problem then 
depends only on the off-diagonal operators $\hat H_{\rm off}$. They
are associated with various constraints, and cannot be inserted or removed at 
a single position only. They can always be inserted and removed pairwise. 
One way to do this is in substitutions with diagonal operators, according to
\begin{equation}
\hat H_{\rm dia},\hat H_{\rm dia} \leftrightarrow 
\hat H_{\rm off},\hat H^\dagger_{\rm off}.
\label{pair}
\end{equation}
In some one-dimensional models, the above types of updates are sufficient for 
achieving ergodicity. In other cases, more complicated updates are also 
required (e.g., involving off-diagonal operators forming loops around 
plaquettes in 2D). The constraints and weight changes associated with local 
updates involve only operators present in $S_L$ which act on a small number 
of lattice sites surrounding those directly affected by the update. Typically,
this allows for a sampling scheme for which the computation time scales as 
$N\beta$.\cite{sse2}

\section{Relation to the Perturbation expansion}

In this Section we discuss the general principles of carrying out importance 
sampling of the standard perturbation expansion in the interaction 
representation. This starting point for a QMC scheme was recently suggested 
by Prokof`ev {\it et al.}.\cite{prokofev} We here show that the 
configuration space of this method is closely related to that of the SSE 
method. We also derive expressions for several types of observables.

The partition function for a Hamiltonian
\begin{equation}
\hat H = \hat D + \hat V ,
\end{equation}
with a diagonal (unperturbed) part $\hat D$ and an off diagonal (perturbing) 
part $\hat V$ is given by the standard time-ordered perturbation expansion 
in $\hat V$,
\begin{equation}
Z= \sum\limits_{n=0}^\infty  (-1)^n
\int_0^\beta d\tau_1 \int_0^{\tau_1} d\tau_2
\cdots \int_0^{\tau_{n-1}} d\tau_n 
{\rm Tr} \bigl\lbrace \hat V(\tau_1) V(\tau_2) \cdots V(\tau_n) \bigr\rbrace,
\end{equation}
where the time dependence in the interaction representation is
$V(\tau)={\rm e}^{\tau \hat D} V {\rm e}^{-\tau \hat D}$. In the same
way as was done in the SSE scheme, $\hat V$ can be decomposed into 
operators that satisfy the requirement (\ref{requirement}), now in the basis 
$\lbrace |\alpha \rangle\rbrace$ where $\hat D$ is diagonal:
\begin{equation}
\hat V = \sum\limits_{b=1}^{M_V} \hat H_b .
\label{sumv}
\end{equation}
For a given model,
the operators in the above sum are of course a subset of those in the
SSE Hamiltonian (\ref{hsum}), where we now define the indexing such that all 
$\hat H_b$ with $b > M_V$ are diagonal. An index sequence defining a product 
of $n$ of the operators $\hat H_b$ is defined as before. In order to 
distinguish the SSE sequence $S_n$, which contains off diagonal as well as 
diagonal operators, from the perturbation expansion sequence containing only 
off-diagonal operators, we denote the latter by $T_n$:
\begin{equation}
T_n = (b_1,\ldots,b_n),\quad b_p \in \lbrace 1,\ldots ,M_V \rbrace .
\end{equation}

Expanding the trace over diagonal matrix elements gives
\begin{equation}
Z=  \sum\limits_{\alpha} \sum\limits_{n=0}^\infty \sum\limits_{T_n}
\int_0^\beta d\tau_1 \int_0^{\tau_1} d\tau_2
\cdots \int_0^{\tau_{n-1}} d\tau_n W(\alpha ,T_n,\lbrace \tau
\rbrace),
\label{irpartition}
\end{equation}
where $\lbrace\tau\rbrace$ is a short-hand for the set of times 
$\lbrace \tau_1,\ldots ,\tau_n\rbrace$. The weight is
\begin{equation}
W(\alpha ,T_n,\lbrace\tau\rbrace) = 
(-1)^n \Bigl ({\rm e}^{-\beta E_0 }\prod\limits_{p=1}^n
{\rm e}^{-\tau_p (E_{p}-E_{p-1})} \Bigr ) \Bigl \langle \alpha \Bigl |
\prod\limits_{p=1}^n \hat H_{b_p} \Bigr | \alpha \Bigr \rangle ,
\label{wtau}
\end{equation}
where $E_p= \langle\alpha (p)|\hat D|\alpha (p)\rangle$.

Now, consider an SSE index sequence $S_n=(b_1,\ldots ,b_n)$, containing $m$ 
indices $b_p \le M_V$, corresponding to $m$
off-diagonal and $n-m$ diagonal operators. Removing all the indices $b_p > 
M_V$ from $S_n$ results in a valid sequence $T_m$. We use the notation
$[S_n]$ for this ``projection'' of $S_n$ onto the corresponding $T_m$; 
$[S_n] = T_m$. Since there are no convergence issues for a finite lattice 
model at finite $\beta$, neither for SSE nor for the perturbation expansion,
the weights of the two formulations must be related according to
\begin{equation}
\sum\limits_{n = m}^\infty
\sum\limits_{[S_n] = T_m} W(\alpha ,S_n) =
\int_0^\beta d\tau_1 \int_0^{\tau_1} d\tau_2
\cdots \int_0^{\tau_{m-1}} d\tau_m W(\alpha ,T_m,\lbrace \tau\rbrace).
\label{wrelation}
\end{equation}
Hence, for a given sequence of off-diagonal operators, the time 
integrals of the perturbation expansion correspond to a summation over
all possible augmentations with diagonal operators in the SSE scheme. The 
equality (\ref{wrelation}) can be explicitly verified in the extreme case
where there are no diagonal operators in $\hat H$, i.e., $\hat D=0$ (for 
example, the XY-model in the $z$-component basis). In this case 
$[S_n]=S_n=T_n$, and the $\tau$-integrals above give $\beta^n/n!$, which 
is exactly the prefactor of the SSE weight (\ref{wn}). 

For a nonzero $\hat D$, the dominant SSE strings contain a finite fraction 
of diagonal operators, and the integrals in the interaction representation 
become nontrivial. In constructing a simulation algorithm based on one
of these expansions, one hence has to weight the disadvantage of a longer 
operator string in the SSE scheme against a more complicated weight function 
in the case of the perturbation expansion. Although the perturbation
series integrand is formally simple, it does not appear to be feasible to 
carry out the time integrals analytically. It is, however, straight-forward 
to include an importance sampling of the times in the Monte Carlo procedures. 
It is likely that sampling the perturbation expansion will be more efficient 
than the SSE algorithm in cases where the diagonal term dominates. The 
average length of the perturbation expansion is then significantly shorter 
than the SSE string. Note, however, that the average length of the 
perturbation expansion has the same scaling $\sim\beta N$ as the SSE 
string length, as will be discussed further below.

In constructing a function $A$ (the estimator) measuring an operator $\hat A$ 
on the configuration space, symmetries of the space should be taken into 
account in order to reduce the statistical fluctuations. An evident one is 
the translational symmetry of periodic (non-random) lattices. Another one is 
the periodicity in the imaginary time (or SSE propagation) direction, 
originating from the cyclic property of the trace operation. In the SSE 
scheme, this is manifested by
\begin{equation}
W(\alpha,S_n) = W(\alpha (p),S_n[p]), \quad p=0,\ldots ,n,
\label{cyclic1}
\end{equation}
where $S_n[p]$ is the index sequence obtained by $p$ times cyclically 
permuting $S_n$, and $\alpha (p)$ refers to the $p$ times propagated
state (\ref{alphap}). One can therefore average the measurements over all
$p=0,\ldots ,n$, an example of which is seen in Eq. (\ref{diacorr}).

The perturbation expansion involves the imaginary times only in 
the form of differences. The part of the weight (\ref{wtau}) containing 
the times can be rewritten as
\begin{equation}
{\rm e}^{-\beta E_0} \prod\limits_{p=1}^n
{\rm e}^{-\tau_p (E_{p}-E_{p-1})} =
\prod\limits_{p=1}^{n} {\rm e}^{-E_p\Delta_{p}} ,
\label{diffexp}
\end{equation}
where $\Delta_p=\tau_{p}-\tau_{p+1} +\sigma\beta$ is the time difference
between the operators at positions $p,p+1$, with $\sigma =0,1$ chosen such 
that $\Delta_p \in [0,\beta]$, and $\tau_n=\tau_0$. Shifting all times by an 
equal amount $\delta$ therefore 
does not change the weight, provided that the shifted times $\tau_i + \delta$ 
obey the limits of the time-ordered integration. Hence, $\delta \in [-\tau_n,
\beta -\tau_1]$. Cyclically permuting $T_n$ and $\lbrace \tau \rbrace$, and
including an appropriate uniform shift $\delta$, is also allowed. We define a 
permutation of the times with an implicit shift, such that the last time in 
the permuted sequence is at its lower bound $0$. Denoting the $p$ times 
permuted set $\lbrace \tau (p) \rbrace$, the zero times permuted 
$\lbrace \tau (0)\rbrace$ hence corresponds to a uniform shift 
$\delta=-\tau_n$. A permutation with an additional shift is denoted 
$\lbrace \tau (p)+\delta\rbrace$. Hence we have
\begin{equation}
W(\alpha,T_n,\lbrace \tau \rbrace ) =
W(\alpha(p),T_n(p),\lbrace \tau(p)+\delta \rbrace ), \quad
\delta \in [0,\Delta_p] .
\label{cyclic2}
\end{equation}

We now derive expressions for some important types of operator 
expectation values within the perturbation expansion scheme, and contrast 
them with those of the SSE formulation. 

First, we consider an equal-time correlation function between two diagonal 
operators, 
\begin{equation}
C_{ij}=\langle \hat A_j\hat A_i \rangle.
\end{equation}
In the SSE approach,
the expansion of Tr$\lbrace \hat A_j\hat A_i {\rm e}^{-\beta\hat H}\rbrace$ 
leads to the same sum as in Eq.~(\ref{ssepartition}), with each term 
multiplied by the 
eigenvalue $a_i[0]a_j[0] = \langle \alpha |\hat A_j\hat A_i |\alpha\rangle$. 
Using the cyclic property (\ref{cyclic1}) then leads to Eq.~(\ref{diacorr}).
In the interaction representation, Eq.~(\ref{cyclic2}) implies that each 
cyclic permutation $p$ should be weighted with the time interval $\Delta_p$.
Since $\sum_p\Delta_p = \beta$ we get
\begin{equation}
C_{ij} = {1\over \beta} \left \langle \sum\limits_{p=1}^n
\Delta_pa_i[p]a_j[p] \right \rangle .
\label{irdiacorr}
\end{equation}
This type of expression is, of course, valid for any diagonal operator.

The SSE expression for the static susceptibility,
\begin{equation}
\chi_{ij} = \int_0^\beta d\tau 
\langle \hat A_j(\tau) \hat A_i(0) \rangle ,
\end{equation}
can be obtained by integrating the time-dependent expectation value
(\ref{taudia}) over $\tau$. Alternatively, one can include a source
$\sum_j h_j\hat A_j$ in the Hamiltonian, and calculate the response function 
via
\begin{equation}
\chi_{ij} = {\partial \langle \hat A_i \rangle \over \partial h_j }
\Bigr |_{h_j=0}.
\label{response}
\end{equation}
The result is \cite{sse1,sse2}
\begin{equation}
\chi_{ij} = \left\langle {\beta\over n(n+1)} 
\left (\sum\limits_{p=0}^{n-1} a_i[p] \right )
\left (\sum\limits_{p=0}^{n-1} a_j[p] \right )
+ {\beta\over (n+1)^2} \sum\limits_{p=0}^{n} a_i[p]a_j[p] \right\rangle .
\end{equation}
In the interaction representation, the derivative (\ref{response})
applied to $\langle \hat A_i \rangle = \sum_p \Delta_p a_j[p]/\beta$ gives
\begin{equation}
\chi_{ij} = {1\over \beta} \left\langle
\left (\sum\limits_{p=1}^{n} \Delta_p a_i[p] \right )
\left (\sum\limits_{p=1}^{n} \Delta_p a_j[p] \right ) \right\rangle.
\end{equation}
Hence, with both methods, there is a simple exact estimator for the static
susceptibility. This is important in view of the fact that the discretization
can introduce spurious temperature dependences in divergent susceptibilities 
calculated using standard worldline methods, due to a combination  of Trotter 
errors and numerical integration errors.\cite{fye} 

A second class of observables easily accessible in SSE as well as real-space 
path integral formulations is one involving the operators 
$\hat H_b$ present in the Hamiltonian. First, consider a single operator
$\langle \hat H_b\rangle$. In the SSE formalism, the estimator is
\begin{equation}
\langle \hat H_b\rangle = -\langle N(b) \rangle /\beta,
\label{sseoff}
\end{equation}
where $N(b)$ is the total number of indices $b_i=b$ in the sequence $S_n$. 
This formula can easily be derived by noting that the expansion of the
numerator Tr$\lbrace {\rm e}^{-\beta\hat H}\hat H_b\rbrace$ leads to a 
one-to-one
correspondence with a subset of the terms in Eq.~(\ref{ssepartition}), 
namely, those for which the last index $b_n=b$. The terms
are related by a factor $-n/\beta$, which hence is the contribution
to $\langle \hat H_b\rangle$ from these partition function configurations.
For the terms with $b_n \not= b$ the contribution is 
zero. Averaging over all cyclic permutations then gives (\ref{sseoff}).
From this result, Eq.~(\ref{energy}) for the SSE internal energy estimator
follows. 

In view of the relation (\ref{wrelation}) between SSE and the perturbation
expansion, we would expect Eq.~(\ref{sseoff}) to be valid also for an 
off-diagonal operator in the interaction representation scheme. Proceeding as 
in the SSE derivation, there is again a one-to-one correspondence between the 
terms of the expansion of Tr$\lbrace{\rm e}^{-\beta\hat H}\hat H_b\rbrace$
and a subset of those in $Z$. However, the situation is complicated by the 
fact that, for a given power $n$, the terms of $Z$ have one more time 
integration. In order to properly relate the terms to each other, we can 
formally introduce another integral,
\begin{equation}
{1\over \tau_{n-1}} \int_0^{\tau_{n-1}} d\tau_{n} = 1,
\label{unitint}
\end{equation}
in the expansion of the numerator.
Terms of order $n-1$ are then in a one-to-one correspondence with terms of 
order $n$ in the partition function (\ref{irpartition}). The lack of the 
time-dependent exponential associated with the last operator $\hat H_b$ 
and the factor $1/\tau_{n-1}$ in (\ref{unitint}) imply a contribution 
$-{\rm e}^{\tau_n(E_{n}-E_{n-1})}/\tau_{n-1}$, if $b_n=b$. 
By the cyclic property (\ref{cyclic2}), this can be averaged over the 
time range $\Delta_p$ for each $p$, giving
\begin{equation}
\langle \hat H_b\rangle = -{1\over \beta}
\left\langle\sum\limits_{p=1}^n I_b(p)K(p)\right\rangle ,
\label{hexp1}
\end{equation}
where $I_b(p)=1$ if $b_p=b$ and $I_b(p)=0$ otherwise. 
Since $\tau_{p-1}=\tau_p+\Delta_{p-1}$, the contribution if $b_p=b$ is
\begin{equation}
K(p)=\int_0^{\Delta_p}
d \tau_p {{\rm e}^{\tau_p(E_{p}-E_{p-1})}\over \tau_{p}+\Delta_{p-1}} =
{\rm e}^{-\Delta_{p-1}(E_{p}-E_{p-1})}
\int_{\Delta_{p-1}}^{\Delta_p+\Delta_{p-1}} 
dx {{\rm e}^{x(E_{p}-E_{p-1})}\over x} .
\label{tintegral}
\end{equation}
This integral cannot be solved in closed form, except if $E_{p}-E_{p-1}=0$. 
We expect (\ref{hexp1}) and (\ref{sseoff}) to be equivalent. Therefore,
the average of (\ref{tintegral}) over all times must give $1$. In fact,
this is the case already for the average over all $\tau_p \in [\tau_{p+1},
\tau_{p-1}]$, which is equivalent to the average over all $\Delta_{p-1}$ 
in the allowed range $[0,\Delta_{p-1}+\Delta_p]$, with
$\Delta_{p-1}+\Delta_p$ kept constant. In doing this averaging, the integral
(\ref{tintegral}) has to be weighted by the relative probability of a given 
$\Delta_{p-1}$, which according to Eq.~(\ref{diffexp}) is 
$\sim {\rm e}^{\Delta_{p-1}(E_{p}-E_{p-1})}$. The resulting double
integral can be solved, with the result 
\begin{equation}
K(p) =
{\int_0^{\Delta_{p-1}+\Delta_p} dy 
\int_{y}^{\Delta_{p-1}+\Delta_p}
dx {1\over x} {\rm e}^{x(E_{p}-E_{p-1})} \over 
\int_{0}^{\Delta_{p-1}+\Delta_p} dx {\rm e}^{x(E_{p}-E_{p-1})}} = 1.
\end{equation} 
Thus, we have shown that (\ref{hexp1}) indeed reduces to the SSE estimator 
(\ref{sseoff}). This then also implies that the average length of the 
perturbation expansion is given by $\langle n\rangle = \beta |\langle \hat V 
\rangle |$, which scales as $\beta N$. 

In order to derive an expression for an off-diagonal equal-time correlation 
function of the type
\begin{equation}
F(b_1,b_2) = \langle \hat H_{b_1} H_{b_2} \rangle ,
\end{equation}
one can proceed along the same lines as for the single operator considered
above. The SSE expression is again formally
very simple. Each occurrence in $S_n$ of a pair of indices $b_1b_2$ 
gives a constant contribution. Denoting by $N(b_1b_2)$ the number of such
pairs of adjacent operators, the result is \cite{sse2}
\begin{equation}
F(b_1,b_2) = \langle (n-1)N(b_1b_2) \rangle / \beta^2 .
\label{sseoffcorr}
\end{equation}
In the interaction representation formalism, the estimator is
\begin{equation}
F(b_1,b_2) = \left\langle \sum\limits_{p=1}^n I_{b_1b_2}(p-1,p)K(p-1,p)
\right\rangle,
\label{offcorr}
\end{equation} 
where $I_{b_1b_2}(p-1,p)=1$ if the indices at the adjacent positions $p-1,p$ 
are $b_1$ and $b_2$, and zero otherwise. In order to evaluate the contribution
$K(n-1,n)$ from a pair at the last two positions, $n-1$ and $n$, we now 
insert a double integral
\begin{equation}
{2\over (\tau_{n-2})^2} \int_0^{\tau_{n-2}} d\tau_{n-1} 
\int_0^{\tau_{n-1}} d\tau_{n} = 1,
\label{unitint2}
\end{equation}
for a term of power $n-2$ in the numerator. Performing the appropriate 
cyclical permutations and time averages analogous to the ones discussed above, 
the contribution from an arbitrary pair of adjacent operators is
\begin{equation}
K(p-1,p)={\int_0^{D_p} dz \int_0^{z} dy 
\int_{z}^{D_p}
dx {1\over x^2} {\rm e}^{x(E_{p}-E_{p-2})} \over
\int_0^{D_p} dy \int_{y}^{D_p} dx
{\rm e}^{y(E_{p}-E_{p-1})} {\rm e}^{x(E_{p-1}-E_{p-2})}} ,
\end{equation} 
where 
\begin{equation}
D_p=\Delta_{p-2}+\Delta_{p-1}+\Delta_{p}.
\end{equation}
Calculating the integrals results in
\begin{equation}
K(p-1,p)= {  (E_{p-1}-E_{p-2})/\beta \over 
1 - {(E_{p}-E_{p-2})({\rm e}^{D_p(E_{p-1}-E_{p-2})}-1) \over
    (E_{p-1}-E_{p-2})({\rm e}^{D_p(E_{p}-E_{p-2})}-1) }},
\end{equation}
where special cases such as $E_{p}-E_{p-2}=0$ should be treated as 
limiting values. In order to relate this much more complicated expression to 
the simple SSE result of a constant contribution $(n-1)/\beta^2$
[Eq.~(\ref{sseoffcorr})], we note that
a typical value of the time interval $D_p$ is $3\beta/n$, and 
$K(p-1,p) \sim 1/(\beta D_p)$ for $D_p$ small. Hence, a typical value of 
$K(p-1,p)$ is $\sim n /\beta^2$, i.e. of the same order as the SSE 
contribution.

The SSE estimator for the static off-diagonal susceptibility,
\begin{equation}
\bar\chi_{ij} = \int_0^\beta d\tau 
\langle \hat H_{b_2}(\tau) \hat H_{b_1}(0) \rangle ,
\end{equation}
is given by the remarkably simple formula \cite{sse2}
\begin{equation}
\bar\chi_{ij} = \left \langle N(b_1)N(b_2)
- \delta_{b_1,b_2}N(b_1)\right\rangle /\beta .
\label{offsusc}
\end{equation}
As this expression 
only involves counting the numbers of indices $b_1$ and $b_2$ in the sequence,
it must [by the configuration relation, Eq.~(\ref{wrelation})] be the correct 
expression in the interaction representation as well [in contrast, the 
equal-time correlation function discussed above involves pairs of indices, the 
distributions of which are different in SSE and the interaction representation,
due to the diagonal operators present in SSE]. We shall not prove this
explicitly here.

The above derivations have clearly shown the close relationships 
between the SSE configuration space and the continuous time path integral.
We note that in cases where the expressions differ, they are generally
formally simpler in the SSE case. In this sense, the SSE propagation
dimension is a more natural representation of the quantum fluctuations 
than imaginary time.

At this stage a reader may wonder why the expansion is dominated by such 
large powers $\langle n\rangle \sim N\beta$, independent of the size of the 
perturbation, and why is it not dominated by small powers when the 
perturbation theory converges for an infinite system at $T=0$. The answer 
lies in the fact that the stochastic sampling is done for the partition 
function, for which there is never a convergent expansion as $T\to 0$ 
and the size of the system goes to infinity. To clarify the situation, 
let us assume that the off-diagonal part of the Hamiltonian is multiplied 
by a perturbation parameter $\lambda$. Furthermore, the free energy per 
unit volume, $f$, has a convergent expansion 
\begin{equation}
f(\lambda)= f_0+f_1\lambda + O(\lambda^2).
\end{equation}
Then the series expansion for the partition function for a system of volume
$N$  becomes
\begin{equation}
Z=e^{-\beta N f}\propto e^{-\beta N f_1}=\sum_n a_n\lambda^n,
\end{equation}
with $|a_n|=(\beta N)^n |f_1|^n/(n)!$. It is easy to show that $a_n$
is maximum for $n=\beta N |f_1|$ in agreement with Eq.~(\ref{sseoff}).

An updating scheme for importance sampling of the perturbation series can 
now be constructed along the lines discussed in the previous section in
the context of the SSE method. The differences are only in the weight 
function, which involves a set of times which also is sampled stochastically.
As a pedagogical example, in the following section we develop the details of 
an algorithm for the simple case of the anisotropic Heisenberg chain. In
Sec.~V we extend the scheme to include couplings to phonon degrees of
freedom (spin-Peierls model).

\section{Algorithm for the Heisenberg Chain}

Here we describe the details of a perturbation series algorithm developed 
for the anisotropic $S=1/2$ Heisenberg chain. We discuss some properties of 
the method and use exact diagonalization results for small systems as well
as known analytical results for the thermodynamic limit to show that very 
accurate, unbiased results can indeed be produced.

\subsection{Construction of the Algorithm}

The model we consider here is defined by the Hamiltonian
\begin{equation}
\hat H = J\sum\limits_{i=1}^N \bigl [ S^z_iS^z_{i+1} +
\hbox{$\Delta\over 2$} (S^+_iS^-_{i+1} + S^+_{i+1}S^-_{i}) \bigr ],
\quad (J > 0),
\label{hheisenberg}
\end{equation}
with periodic boundary conditions. $\Delta$ controls the anisotropy, with
$\Delta=1$ corresponding to the isotropic Heisenberg point.

We wish to construct an algorithm in which, as in the SSE scheme, Monte Carlo 
updates that change the expansion order, $n$, are accomplished by inserting 
or removing diagonal operators one at a time, and off-diagonal operators are 
inserted or removed pairwise in substitutions with diagonal operators.
Since there are only off-diagonal 
operators in the perturbation expansion string, we add constants to the 
Hamiltonian, and formally consider these as part of the perturbation. For
the spin chain at hand we define
\begin{mathletters}
\begin{eqnarray}
\hat H_{1,b} & = & -I \\
\hat H_{2,b} & = & S^+_iS^-_{i+1} + S^+_{i+1}S^-_{i},
\end{eqnarray}
\end{mathletters}
and write the perturbation as
\begin{equation}
\hat V = \Bigl ( {\Delta\over 2} \Bigr )
\sum\limits_{a=1}^2 \sum\limits_{b=1}^N \hat H_{a,b}.
\end{equation}
For $N$ even, only operator strings with an even number of the off-diagonal 
operators $\hat H_{2,b}$ contribute to the partition function. The weight 
(\ref{wtau}) is hence positive definite. The matrix element of an allowed 
operator string equals one, and therefore
\begin{equation}
W(\alpha,T_n,\lbrace \tau \rbrace )=
(\Delta /2)^n {\rm e}^{-\beta E_0} \prod\limits_{p=1}^n
{\rm e}^{-\tau_p (E_{p}-E_{p-1})}.
\label{weight}
\end{equation}
$T_n$ is now for convenience defined as a sequence of index pairs 
\begin{equation}
T_n = [a_1,b_1],[a_2,b_2],\ldots,[a_n,b_n],
\end{equation}
with $a_p  \in \lbrace 1,2\rbrace$ and $b_p \in \lbrace 1,\ldots ,N \rbrace$
referring to the operator type and lattice bond (nearest-neighbor spin
pair), respectively. Using the fixed-length scheme developed for the SSE 
algorithm, we define $\hat H_{0,0}=I$, and insert $L-n$ of these in each 
string. Taking into account the number of possible insertions then gives
\begin{equation}
W(\alpha,T_L,\lbrace \tau \rbrace )= 
{(\Delta /2)^n (L-n)!n!\over L!}
{\rm e}^{-\beta E_0} \prod\limits_{p=1}^n
{\rm e}^{-\tau_p (E_{p}-E_{p-1})},
\label{lweight}
\end{equation}
where $n$ is the number of non-$[0,0]$ elements in $T_L$. There are no
times associated with the augmentation operators $H_{0,0}$, and the index $p$
in the product therefore refers to the $p$th non-$[0,0]$ operator. 

The Monte Carlo sampling is based on updates of the types (\ref{single})
and (\ref{pair}). Using $[a,b]_p$ as an alternative to the notation
$[a_p,b_p]$, the update changing the power $n$ by $\pm 1$ is
\begin{equation}
[0,0]_p \leftrightarrow [1,b]_p,
\label{upd1}
\end{equation}
and the simplest update involving off-diagonal operators is
\begin{equation}
[1,b]_{p_1},[1,b]_{p_2}  \leftrightarrow [2,b]_{p_1},[2,b]_{p_2} .
\label{upd2}
\end{equation}
These local updates are sufficient for generating all operator strings for
an open chain, or a periodic system in the zero winding number sector. An 
update changing the winding number will
be discussed further below. For a simulation in the canonical ensemble, i.e.,
with the total  magnetization $m = \sum_i S^z_i$ fixed, no further updates
of the state $|\alpha \rangle$ are required, since (\ref{upd2}) also implies
flips of nearest-neighbor spins in the propagated states $|\alpha (p)\rangle$
with $p = p_1,\ldots ,p_2-1$ (here and in the following, the periodicity of
the sequence is always implied, so that if $p_1 > p_2$, the affected states 
are $p_1,\ldots,L-1,0,\ldots,p_2-1$). In the grand canonical ensemble, global
spin flips changing the magnetization of all the propagated states also have 
to be carried out.

Fig.~\ref{conf} shows a graphical representation of a configuration
generated for an 8-site isotropic Heisenberg chain. This type of representation
emphasizes that in this simulation scheme the ordered sequence of operators 
is the central element, and the times formally can be thought of as
auxilliary variables associated with the operator positions. In the 
simulation, only one of the states $|\alpha (p)\rangle$ needs to be stored, 
since all the other ones are uniquely defined given the operator sequence 
and can be generated as needed.

First, consider the single-operator update (\ref{upd1}). This can be 
attempted consecutively at all positions $p=1,\ldots,L$ for which $a_p 
\in {\lbrace 0,1 \rbrace}$. In calculating the Metropolis acceptance 
probabilities for such updates, the fact that the augmentation operators 
$\hat H_{0,0}$ are not associated with time integrals has to be taken
into account. We now define the time difference $\bar\Delta_p = 
\tau{(<p)}-\tau{(\ge p)}$, where $\tau{(<p)}$, and $\tau{(\ge p)}$ are the
times associated with the non-$[0,0]$ operator closest to $p$, with
position indices $< p$ and $\ge p$, respectively. In a substitution
$[0,0]_p \rightarrow [1,b]_p$, $b$ is chosen at random, and a time
$\tau_p$ in the range $[\tau{(>p)},\tau{(< p)}]$ is generated. In the 
direction $[1,b]_p \rightarrow [0,0]_p$, the only action is to replace 
$[1,b]$ with $[0,0]$ and discard the time $\tau(p)$. One can easily 
verify that detailed balance with the distribution (\ref{lweight}) is 
maintained with the following acceptance probabilities (note that the 
energy difference $E_p-E_{p-1}=0$ for $[a_p,b_p] = [1,b]$):
\begin{mathletters}
\begin{eqnarray}
P\left ([0,0]_p \to [1,b]_p \right ) & = & {\rm min}
\left [{(\Delta/2)\bar\Delta_p N(n+1) \over L-n},1 \right ] \\
P\left ( [1,b]_p \to [0,0]_p \right ) & = & {\rm min}
\left [{L-n+1\over (\Delta/2)(\bar\Delta_p+\bar\Delta_{p+1})Nn },1 \right ].
\end{eqnarray}
\label{accdia}
\end{mathletters}

The pair substitutions (\ref{upd2}) are associated with constraints. 
In the $\rightarrow$ direction, the first requirement is that $S^z_{b}[p_1-1]=
-S^z_{b+1}[p_1-1]$. In either direction, an update flips the spins 
$S^z_b[p]$ and $S^z_{b+1}[p]$ in the states $|\alpha (p)\rangle$ with $p_1 
\le p < p_2$. This implies that there may be no operators $[2,b-1]$ or 
$[2,b+1]$ present between $p_1$ and $p_2$, and hence that $S^z_{b+1}[p]=-
S^z_{b+1}[p]$ for all $p$ in the range affected by the update. The change of 
the weight (\ref{lweight}) in an allowed update at $b$ then depends on the 
energy differences $E_p-E_{p-1}$ in the local 4-spin substates 
$|S^z_{b-1},S^z_{b},S^z_{b+1},S^z_{b+2}\rangle _p$ for $p_1 \le p \le p_2$, 
but only for those $p$ with operators acting on the spins of this substate.

The locality of the constraints and the weight changes allows for a fast 
updating carried out on {\it subsequences}.\cite{sse2} A subsequence 
$b$ contains all operators $[2,b]$ present in $T_L$, and those $[1,b]$
operators appearing between antiparallel spins $S^z_b$ and $S^z_{b+1}$. 
Information on the constraints imposed by the 
presence of the nearest-neighbor operators $[2,b-1]$ and $[2,b+1]$ is also
part of the subsequence. Operators $[2,b-2]$ and $[2,b+2]$ do not 
impose constraints, but affect the edge spins of the substates 
$|S^z_{b-1},S^z_{b},S^z_{b+1},S^z_{b+2}\rangle$, and therefore also have to 
be included in the subsequence, so that the acceptance probabilities can be 
calculated. All the 4-spin substates acted upon by the operators of the 
subsequences are also stored for this reason. Clearly, subsequences $b$ 
and $b'$ can be updated independently of each other if 
$|b'-b| >2$. Since we normally study chains with $N$ a multiple of $4$, we 
simultaneously construct the subsequences for all bonds separated by 3 other 
bonds, and update these one by one. Four such partition--updating cycles are 
then needed for updating all the bonds. 

The subsequence information is arranged as follows: The length of subsequence 
$b$ is denoted $L_b$, and is the number of operators $[1,b]$, $[2,b]$, 
$[2,b-2]$, and $[2,b+2]$ present in $T_L$. These operators are represented by 
the integers $1-4$, and are stored in lists $A_b(1,\ldots,L_b)$. The original 
positions of these operators in $T_L$ are needed for re-merging the updated 
subsequences into an updated full sequence, and are stored in lists 
$P_b(1,\ldots,L_b)$. The constraining operators $[2,b \pm 1]$ do not have 
to be stored. Instead, lists $F_b(1,\ldots,n_b)$ are created, such that 
$F_b(i)=1$ if there are constraining operators (one or several) in $T_L$ 
between positions $P_b(i)$ and $P_b(i+1)$, and $F_b(i)=0$ otherwise. The 
4-spin substates are encoded as single integers ($1-16$), and stored in 
lists $S_b(1,\ldots,L_b)$.  

For updating the subsequences, we use the scheme introduced in
Ref.~\onlinecite{sse2} (other methods are also possible). An attempt to carry 
out a substitution (\ref{upd2}) in a given subsequence $b$ consists of the 
following steps. A position $i_1$ such that $F_b(i_1)=0$ is first chosen at 
random. One then searches in the forward direction for the first position $j$ 
for which $A_b(j)=A_b(i_1)$ and $F_b(j)=1$. This position is the one furthest
away from $i_1$ which can be considered, together with $i_1$, in a pair
substitution. Note that position $i=1$ follows $i=L_b$ due to the periodicity,
and the search is terminated at $i=i_1-1$ if this position is reached (and 
then $j=i_1-1$). During the search, the positions $i$ of all encountered 
operators $A_b(i)=A_b(i_1)$ are stored. One of these, $i=i_2$, is then 
selected at random, and the pair $\lbrace A_b(i_1),A_b(i_2)\rbrace$ is
replaced by $\lbrace A'_b(i_1),A'_b(i_2) \rbrace = 
\lbrace 2-A_b(i_1),2-A_b(i_2) \rbrace$ with a probability satisfying
detailed balance. 

The total probability of making a certain pair-substitution is the product 
of the probability $P_{\rm select}[A(i_1),A(i_2)]$ of selecting the operators 
at positions $i_1$ and $i_2$, and the acceptance probability 
$P_{\rm accept}[A(i_1)A(i_2) \rightarrow A'(i_1)A'(i_2)]$. 
One can show that the selection probabilities with the above
procedures are the same in both directions,\cite{sse2} i.e., 
$P_{\rm select}[A(i_1),A(i_2)]=P_{\rm select}[A'(i_1),A'(i_2)]$, 
if the attempt is {\it cancelled} with a probability
\begin{equation}
P_{\rm cancel}[AA \rightarrow A'A'] = 1-{N(A) \over N(A) + N(A')},
\label{attlink}
\end{equation}
where $N(A)$ and $N(A')$ are the numbers of operators $A$ and $A'$ found 
between $i=i_1$ and $i=j$ in the search [excluding the first operator 
$A_b(i_1)=A$]. If the attempt is not cancelled at this stage, a final 
acceptance probability is calculated on the basis of the weight 
(\ref{lweight}), using the
stored substates, and the corresponding states modified due to
the replaced operators (obtained by propagating with the updated subsequence 
segment). The Metropolis acceptance probability is
\begin{equation}
P_{\rm accept}[AA \rightarrow A'A'] = 
{\rm min}\Bigl [ {\rm e}^{-\beta (E'_0-E_0)} \prod\limits_{i={p_1}}^{p_2}
{\rm e}^{-\tau_{P_b(i)} (E'_{i}-E'_{i-1}+E_{i}-E_{i-1})},1\Bigr ],
\end{equation}
where $E_i$, and $E'_i$ are the eigenvalues of $\hat D$ calculated on the 
substate $|S^z_{b-1},S^z_{b},S^z_{b+1},S^z_{b+2}\rangle_i$ before and 
after the operator substitution. Typically, a constraint is 
encountered already a few steps from the starting position $i_1$, and 
therefore the number of operations required per step is rather small
(the cancellation probability (\ref{attlink}) is often $0$).

For each subsequence, a number of updating attempts proportional to the 
number of operators in the subsequence is carried out. The average length of 
the subsequences, and hence the number of operations needed for its updating, 
is proportional to $\beta$ at low temperatures. After updating all the 
subsequences belonging to one out of the four partitions, the updated 
operators are inserted in the full index sequence, and changes in the local 
4-spin substates are copied into the stored full-system state 
$|\alpha \rangle$. The same procedures are repeated for all four 
partitions.

In a periodic system, configurations with a non-zero winding number are
possible, and cannot be generated by the local updates discussed above.
A winding number corresponds to an excess of spin flips in one
direction in the course of the propagation with the operator string, i.e.,
a cyclic permutation of same spins in $|\alpha (L) \rangle$ with respect
to those in $|\alpha (0) \rangle$. The winding number can be changed by
substituting a ``half-ring'' of off-diagonal operators by the complementary
half-ring according to\cite{sse1}
\begin{equation}
[2,b_1]_{p_1},[2,b_2]_{p_2},\ldots [2,b_{N/2}]_{p_{N/2}}  
\leftrightarrow 
[2,b'_1]_{p_1},[2,b'_2]_{p_2},\ldots [2,b'_{N/2}]_{p_{N/2}} ,
\label{ringupd}
\end{equation}
where the bonds $b_1,\ldots,b_{N/2},b'_1,\ldots,b'_{N/2}$ are a permutation
of all the bonds of the periodic lattice. The acceptance rate for this type 
of update decreases rapidly with increasing system size, due to the increasing
number of constraints. In practice, simulations for systems larger than 
$N \approx 16-20$ must be restricted to the sector with zero winding number
The resulting small error is a boundary effect, and vanishes in the 
thermodynamic limit.\cite{phenelius}

At high temperatures, simulations can be carried out in the grand canonical 
ensemble, by including updates of the total magnetization. This is in
principle easily achieved by flipping ``straight lines'' of spins, 
$S^z_i[0],\ldots, S^z_i[L-1]$, which is allowed provided that there are 
no operators $[2,i-1]$ or $[2,i]$ present in $T_L$, and the acceptance 
probability then depends only on the neighbor spins $S^z_{i-1}[0]$ and 
$S^z_{i+1}[0]$. However, the likelihood of an allowed spin flip decreases 
rapidly with decreasing temperature, and in practice simulations for
$T \alt J/10$ have to be carried out in the canonical ensemble.

Finally, we also perform updates of the times $\lbrace \tau \rbrace$ without
changes either in the operator string or the states. A single time $\tau_p$ 
can be updated by generating a new time in the allowed range 
$[\tau_{p+1},\tau_{p-1}]$ (and $\tau_1 \le \beta$, $\tau_n \ge 0$), and 
accepting this with a Metropolis acceptance probability calculated from 
Eq.~(\ref{lweight}):
\begin{equation}
P(\tau_p \to \tau'_p) = {\rm min} \left [ {\rm exp}
[(\tau'_p-\tau_p)(E_{p-1}-E_p)],1 \right ].\label{tupd1}
\end{equation}
The typical time separation, and hence the difference $\tau'_p-\tau_p$, 
scales as $1/N$, and is typically very small. The acceptance rate for these
single-time updates is therefore close to $100\%$ in most cases. It is clear 
that the rate of evolution  of $\lbrace \tau \rbrace$ updated this way 
[in addition to the generation of a random time when inserting an operator 
in an update (\ref{upd1})], will be very slow for large systems. Therefore, 
we consider simultaneously updating a whole set of times 
$\lbrace \tau_{p_1},\ldots \tau_{p_2} \rbrace$, 
using the following scheme.

A position $p_1$ is first chosen at random, and $p_2$ is
chosen as the smaller of $p_1+n_\tau$ and $n$, with $n_\tau$ a number
chosen randomly between 1 and some upper bound $m_\tau$, and $m_\tau$ is
adjusted so that a reasonable acceptance rate ($\approx$ 50\%) is 
maintained. If the weight would be independent of the times, the distribution 
of the times would be uniform within the limits of the time ordered integral.
If the separation $\tau_{p_2} - \tau_{p_1}$ is not too large, the true
distribution will be close to uniform. We therefore attempt to replace the 
selected set of times by a randomly generated ordered set 
$\lbrace \tau'_{p_1},\ldots \tau'_{p_2} \rbrace$, 
with $\tau'_{p_1} \le \tau_{p_1-1}$ ($\tau'_{p_1} \le \beta$ if $p_1=1$)
and $\tau'_{p_2} \ge \tau_{p_2+1}$ ($\tau'_{p_2} \ge 0$ for $p_2 = n$). 
The Metropolis acceptance probability for this multi-time update is
\begin{equation}
P( \lbrace \tau_p\rbrace  \to \lbrace \tau' \rbrace ) = 
{\rm min}\left [ {\rm exp} \left ( \sum_{p_1}^{p_2} 
(\tau'_p-\tau_p)(E_{p-1}-E_p) \right ),1 \right ].
\label{tupd2}
\end{equation}
It is clear that the acceptance rate is essentially determined by the time 
difference $\tau_{p_2}-\tau_{p_1}$, independently of the system size. In
simulations of large systems, the maximum number of simultaneously updated 
times, $m_\tau +1$, can therefore be as high as $10^3$ or higher (in many 
cases, all times can be updated simultaneously). The importance of 
the multi-time updates will be further discussed below.

We define a Monte Carlo step (MC step) as a sequence of diagonal updates
(\ref{upd1}) at all positions in $T_L$, followed by off-diagonal pair updates 
(\ref{upd2}) at all bonds. In a grand canonical simulation, a global
flip of each spin is also attempted, and for simulations of small systems 
with fluctuating winding number ``ring updates'' (\ref{ringupd}) are carried 
out. The number of multi-time updates per MC step is chosen such that, on 
the average, $\approx$50\% of the times are changed. As already noted, an 
MC step requires of the order of $N\beta$ operations. 

\subsection{Performance Tests}

We now present some tests of the accuracy of the method. We also briefly 
address the issues of autocorrelation times, and the equilibration of the 
simulation. We only consider the spin isotropic Heisenberg case [$\Delta=1$ in 
Eq.~(\ref{hheisenberg})]. Detailed numerical finite-temperature results 
for this model have recently been obtained using the SSE method and 
high-temperature expansions.\cite{sss}

In order to verify that the new QMC algorithm indeed produces results free from
detectable systematic errors, we carried out a long simulation of a 12-site
system at inverse temperature $\beta=8$, in the grand canonical ensemble
and with fluctuating winding numbers. These results can be directly compared
with exact diagonalization data.

A useful internal check of the simulation in the isotropic case is the internal
energy calculated in two different ways; from the diagonal nearest-neighbor
correlation function as $E_1 = 3 \langle S^z_i S^z_{i+1}\rangle$, according 
to Eq.~(\ref{diacorr}), and from the expectation value of the off-diagonal
operators, $E_2 = -(3/2)\langle \hat H_{2,i}\rangle$, according to
Eq.~(\ref{sseoff}). For the 12-site system at $\beta=8$, a simulation
consisting of $2 \times 10^8$ MC steps gave $E_1 -0.44372(3)$ and 
$E_2=-0.44368(2)$, where the numbers within parentheses indicate the 
statistical errors. The exact result is $E=-0.443697$. Hence, both QMC 
estimates are accurate to within relative statistical errors of less than 
$10^{-4}$.

We also calculated the static structure factor
\begin{equation}
S(q) = {1\over N} \sum\limits_{j,l} {\rm e}^{-i(j-l)q}
\langle S^z_{j}S^z_l \rangle ,
\end{equation}
and the corresponding static susceptibility
\begin{equation}
\chi(q) = {1\over N} \sum\limits_{j,l} {\rm e}^{-i(j-l)q}
\int_0^\beta d\tau \langle S^z_{j}(\tau)S^z_l (0) \rangle .
\label{statsusc}
\end{equation}
Comparisons with the exact results are presented in Table \ref{tab1}. Here 
the relative accuracy is the highest, $\approx 10^{-4}$, close to 
$q=\pi/2$. The accuracy is lower close to $q=\pi$, due to the strong 
antiferromagnetic fluctuations present in the model. At $q=0$ the accuracy 
is hampered by the low acceptance rate for the spin flips required in the 
grand canonical ensemble.

The above results clearly demonstrate that the method in principle is very 
accurate. We now consider a significantly larger system, and address some
practical issues that arise in realistic simulations.

For the Heisenberg chain, a number of exact results are known in the
thermodynamic limit. For example, Eggert {\it et al.} recently calculated the 
temperature dependence of the uniform magnetic susceptibility $\chi = \chi 
(q \to 0)$, using the thermal Bethe Ansatz.\cite{eggert} We here compare their 
infinite-size result with QMC data for a system with 128 spins. The 
simulations consisted of $3-5 \times 10^7$ MC steps for each temperature. At 
high temperatures ($T/J \ge 0.1$) the grand canonical ensemble was used, and 
at lower $T$ the magnetization was kept fixed at $m^z=0$. In the latter
case, we define the uniform susceptibility as $\chi (q_1)$ where $q_1$
is the smallest non-zero wavenumber; $q_1=2\pi/L$. Fig.~\ref{suscfig} shows 
the results. The relative statistical errors are $\approx 10^{-3}$ down to 
about $T/J=0.15$, below which the accuracy diminishes due to the low 
acceptance rate for the grand canonical spin flips. More accurate results 
could have been obtained by extrapolating to zero wavenumber on the basis of 
a few points close to $q=0$, instead of just using $q=0$. For the grand
canonical simulations, the agreement with the exact result is clearly very 
good, indicating no detectable effects of finite size for $N=128$ at the 
temperatures considered. For the canonical ensemble, all results are
slightly below than the exact curve, suggesting larger finite-size effects 
when the magnetization is not allowed to fluctuate (extrapolating to $q=0$ 
instead of using $q_1$ gives a still larger deviation).

In any Monte Carlo simulation, care has to be taken that the equilibrium
distribution is reached before measurements are taken. In our algorithm,
the cut-off of the perturbation expansion at order $L$ is determined
during the equilibration part of the simulation, by monitoring the power
$n$ and increasing $L$ whenever $n$ exceeds some threshold fraction
of $L$ (typically 90-95\%). We have found that an efficient updating of the
times $\lbrace \tau \rbrace$ is crucial for achieving a rapid equilibration. 
In Fig.~\ref{equifig} we show results for the cut-off versus the simulation 
time for an $N=128$ system at $\beta=8$ for two different ways of updating
the times; consecutively updating all the times one-by-one, according to
Eq.~(\ref{tupd1}), and collectively updating a whole range of times according 
to Eq.~(\ref{tupd2}). With single-time updates, the final cut-off $L=1064$ 
was reached after approximately $20000$ MC steps, whereas the multi-spin 
updating gave the final $L$ already after $288$ steps. In this case, all 
times were updated simultaneously, with an average acceptance rate of 
$\approx 70$\%.

Measuring the expectation value of the constant operators $\hat H_{1,b}$ also
provides for a good check of the equilibration of the sequence length. 
With $J=1$, $-2\langle \hat H_{1,b}\rangle$ should equal $1$, but
before equilibrium is reached a calculation using Eq.~(\ref{sseoff}) 
will deliver a smaller value. In Fig.~\ref{equifig} we also show the 
evolution of this measurement with MC time, for the two different ways of 
updating the times. With multi-time updating, the measured values fluctuate 
around $1$ already after $10^3$ steps, whereas with single-time updates the 
result are too low even after $10^5$ steps. In both cases, subsequent 
simulations consisting of $10^7$ MC steps gave comparable results for all
calculated quantities, with similar statistical errors. Hence, in this case
the multi-time update appears to be important for reaching equilibrium, but 
does not significantly affect the statistics of the simulation otherwise. For 
larger system the single-time update becomes even less efficient, and it is 
likely that the multi-time update then will become important also for the 
statistics of the simulation (i.e., the autocorrelation times of measured 
quantities).

The efficiency of the updating process in generating independent (uncorrelated)
configurations is measured by autocorrelation functions. The optimal frequency
for measuring expectation values on the generated configurations is in 
principle determined by such functions. For a quantity $A$, 
the autocorrelation function is defined as
\begin{equation}
\alpha_A(t) = {\langle A(i+t)A(i)\rangle - \langle A(i)\rangle^2 \over
\langle A(i)^2\rangle - \langle A(i)\rangle^2},
\label{auto}
\end{equation}
where $A(i)$ denotes the value of the simulation estimator after $i$ MC steps,
and the normalization has been chosen such that $\alpha_A (0)=1$. For large 
time separations $t$, one expects $\alpha_A(t)$ to decay as e$^{-t/\xi_A}$, 
where $\xi_\alpha$ is the autocorrelation time. In principle, the long-time 
dynamics of the simulation is characterized by a single autocorrelation time,
i.e., the longest $\xi_A$. An observable of interest may or may not overlap 
with the quantity corresponding to this slowest mode. Hence, in practice one 
will deal with different autocorrelation times for different quantities. 

Successive measurements provide independent information only if they are 
separated by a number of steps sufficiently large for the corresponding 
autocorrelation function to have decayed substantially. In practise, the 
decay may be much faster than e$^{-t/\xi_\alpha}$ for short times, and 
therefore the relevant separation between measurements may be shorter than 
$\xi_\alpha$. In general one expects the autocorrelations for observables
related to the long-distance properties of the model, such as $S(\pi)$ and 
$\chi (\pi)$ for the antiferromagnetic Heisenberg chain,
to decay slower than those for essentially local quantities such 
as the energy, or correlation functions away from the wave number of the 
dominant fluctuations. Fig.~\ref{autofig} shows results for the logarithms 
of the autocorrelation functions of $E$, $S(\pi)$, $\chi (\pi)$, and $\chi(0)$
for a $128$-site system at $\beta=8$. For $S(\pi)$ and $\chi (\pi)$, the 
asymptotic linear decay is the same, as expected, and the autocorrelation
time is $\approx 40$ MC steps. Note, however, tha $\alpha_{S(\pi)}(t)$ 
decays slightly faster than $\alpha_{\chi(\pi)}(t)$ for short times, which 
can be understood on account of $S(\pi)$ 
being an equal time correlation function, whereas $\chi (\pi)$ involves 
also an integration over imaginary time. This effect becomes even more 
pronounced at lower temperatures. For $\chi (0)$ the autocorrelation time is 
considerably longer at this temperature ($\approx 80$ MC steps), due to the 
low acceptance rate for the global spin flips. In contrast, at higher 
temperatures $\chi (0)$ exhibits the shortest autocorrelation time. The 
autocorrelation function for the energy decays very rapidly, and it is 
difficult to see a regime with a linear behavior before statistical noise 
dominates the measurements. One would expect $\alpha_E (t)$ to contain a 
component of the slowest mode of the simulation, but the overlap may be too 
small to detect (and hence the effect on the statistics is essentially 
irrelevant). Clearly, the relevant time scale for measurements of the energy 
is much shorter than its asymptotic autocorrelation time. 

In practice, it is not feasible to measure the autocorrelations in every case. 
As usual, binning the data, so that a single bin represents a simulation time 
much longer than the asymptotic autocorrelation time, ensures proper estimates
of averages and statistical errors. The only concern then is that the time 
spent on the measurements should not dominate the simulation. Measuring 
every $10-20$ MC steps typically only results in a minor overhead.

\section{Spin-Peierls Model}

The effects of phonons on spin systems in one dimension are of great current
interest, in view of the recent discoveries of spin-Peierls transitions in
the inorganic  compounds GeCuO$_{\rm 3}$ (Ref.~\onlinecite{gediscovery}) and 
$\alpha'-$NaV$_{\rm 2}$O$_{\rm 5}$ (Ref.~\onlinecite{nadiscovery}). These 
materials consist of weakly coupled spin-1/2 chains and dimerize at $T=14$ K 
and $36$ K, respectively. A variety of experiments have probed their static 
and dynamic magnetic properties.\cite{geexperiments,naexperiments} 

To date, most numerical work on spin-Peierls models in 1D has focused on 
Lanczos exact diagonalization studies of the ground state of dimerized 
chains, and the finite temperature susceptibility has been calculated using 
complete diagonalization of undimerized chains.\cite{spdiag1,spdiag2,spdiag3} 
The QMC approach discussed in this paper should be an excellent tool for 
detailed non-perturbative studies of spin systems including dynamic, 
i.e., fully quantum mechanical, phonon degrees of freedom.

Early studies of 1D electronic models including phonons were carried out by 
Hirsch and Fradkin, using the worldline QMC method with the phonons 
introduced via their real-space displacement coordinates, in imaginary time 
discretized with the standard Trotter approximation.\cite{hirsch} We have 
developed an extension of the Heisenberg algorithm described in the previous 
Section, with phonons treated in the real-space occupation number basis. We 
believe that this is a more efficient way to include the phonons, and 
furthermore the scheme is exact. The same ideas could also easily be 
implemented within the SSE scheme, but so far we have not done so. The 
present scheme is likely more efficient than SSE at high temperatures, 
where the average phonon occupation number is high and dominates the 
energy. At low temperatures, SSE may in fact be slightly more efficient.

We here restrict ourselves to the perhaps simplest type of spin-phonon
coupling, namely, we consider a dispersionless harmonic oscillator with
frequency $\omega_0$ at each bond $i$ (Einstein phonons). The exchange 
interaction between the spins at sites $i$ and $i+1$ is modulated by the 
phonon displacement $x_i$. The Hamiltonian is hence
\begin{equation}
\hat H = \sum\limits_{i=1}^N (J+\alpha x_i) {\bf S}_i \cdot {\bf S}_{i+1}
+ \omega_0 \sum\limits_{i=1}^N n_i .
\label{sphamiltonian}
\end{equation}
Here $J$ is the ``bare'' exchange, $\alpha$ is the spin-phonon coupling,
and $n_i$ is the phonon occupation number, given in terms of the phonon
creation and destruction operators as $n_i = a^\dagger_ia_i$. We define the
phonon displacement operator as
\begin{equation}
x_i = (a^\dagger_i + a_i)/\sqrt{2} ,
\label{xdef}
\end{equation}
and hence absorb in the coupling $\alpha$ the mass and spring constant of 
the oscillator. The oscillator potential corresponding to the definition 
(\ref{xdef}) is $V(x) = \omega_0x^2/2$.

We remark that the linear coupling is not completely realistic, since it
can lead to negative (ferromagnetic) spin-spin couplings if the fluctuations 
are large. In the low temperature regime this does not occur, and the model 
should then be a good starting point for understanding the effects of 
phonons in real quasi-1D spin systems.

Augier and Poilblanc \cite{augier} have recently studied a model related to 
the Hamiltonian (\ref{sphamiltonian}). They included only the phonons with 
momentum $q=\pi$, which are the ones forming a condensate in the dimerized
state, and carried out $T=0$ Lanczos exact diagonalization in a space with 
a restricted number of these phonons. They found quantitative changes in the
behavior from that of a statically dimerized system. Here we will show that 
the $q \not=\pi$ phonons are also important at finite temperature, 
in particular the $q=0$ ones which lead to a temperature dependent effective 
spin-spin coupling.

\subsection{Algorithm}

In a simple modification of the Heisenberg algorithm described in the
previous Section, we now have additional phonon occupation number operators
in the unperturbed Hamiltonian;
\begin{equation}
\hat D = J_0\sum_{b=1}^N S^z_bS^z_{b+1} + 
\omega_0 \sum\limits_{b=1}^N n_b .
\label{spdiagonal}
\end{equation}
We write the perturbation 
$\hat V = \sum_{a,b}\hat H_{a,b}$ in terms of the following operators: 
\begin{mathletters}
\begin{eqnarray}
\hat H_{1,b} &=& -J_0/2 \\
\hat H_{2,b} &=& (J_0/2)(S^+_{b}S_{b+1}^- + S^+_{b+1}S_{b}^-) \label{h2} \\
\hat H_{3,b} &=& 
(\alpha /2\sqrt{2})(S^+_{b}S_{b+1}^- + S^+_{b+1}S_{b}^-)a^\dagger_b \\
\hat H_{4,b} &=& 
(\alpha /2\sqrt{2})(S^+_{b}S_{b+1}^- + S^+_{b+1}S_{b}^-)a_b \\
\hat H_{5,b} &=& 
(\alpha /\sqrt{2})(S^z_{b}S^z_{b+1}-\hbox{$1\over 4$})a^\dagger_b\\
\hat H_{6,b} &=& 
(\alpha /\sqrt{2})(S^z_{b}S^z_{b+1}-\hbox{$1\over 4$})a_b\\
\hat H_{7,b} &=& 
(\alpha /2\sqrt{2})a^\dagger_b \label{h7} \\
\hat H_{8,b} &=& (\alpha /2\sqrt{2})a_b \label{h8} .
\end{eqnarray}
\end{mathletters}
The definitions of $\hat H_{5,b}$ and $\hat H_{6,b}$ contain off-diagonal
operators $\sim a^\dagger_b$ and $\sim a_b$, which are not present in the 
original Hamiltonian. They have been included in order to make the weight
function positive definite. They also induce a constant shift
in the oscillators, which has no effect other than shifting the value of $J$
by an amount proportional to $\alpha^2$. The operators $\hat H_{7,b}$ and 
$\hat H_{8,b}$ have been included in the Hamiltonian in order to enable 
straight-forward measurements of correlation functions of the displacement 
operators $x_b$, which are off-diagonal in the chosen representation and 
hence must be measured using expressions such as Eqs.~(\ref{offcorr}) and 
(\ref{offsusc}) [the prefactor in (\ref{h7}) and (\ref{h8}) is chosen for 
convenience, but is in principle arbitrary]. Again, these operators just shift 
the effective value $J$ and therefore do not affect the physics of the model. 

In order to compensate for the shift induced by the added operators, we 
define the spin-spin coupling in Eqs.~(\ref{h2}) and (\ref{spdiagonal}) 
with a constant $J_0 \not=J$, determined such that the new bare coupling 
$J=1$ (i.e., its value in the absence of the spin-phonon coupling 
$\alpha\sum_i x_i {\bf S}_i \cdot {\bf S}_{i+1}$, which in turn induces an 
additional shift which {\it is} part of the physics of the model). 
For each bond, the added operators equal $(3\alpha/4)x_b$. 
Thus the shifted oscillator potential is 
$V'(x) = \omega_0(x-3\alpha/{4\omega_0})^2/2$.
The actual bare value of $J$ is then
\begin{equation}
J = J_0 + {3\alpha^2\over 4\omega_0} ,
\end{equation}
and we therefore choose $J_0 = 1- {3\over 4}(\alpha /J)(\alpha /\omega_0)$
and $J=1$. If we do not want to deal with negative values of $J_0$, this 
implies that the maximum coupling constant we can study, for a given 
$\omega_0$, is $\alpha_{\rm max} = 2(J\omega_0/3)^{1/2}$, which is probably 
not a serious restriction for realistic situations. We can relax the 
restriction by choosing
a smaller prefactor in the definition of $\hat H_{7,b}$ and $\hat H_{8,b}$, or
leaving out these operators altogether (which can be done if no phonon
correlation functions need to be measured in the simulation). We note that in 
principle one can use $J_0 < 0$ and avoid the coupling constant restriction. 
However, this causes a sign problem due to frustration. The standard way of 
treating the phonons, in the representation where the displacements $x_i$ are 
diagonal,\cite{hirsch} avoids this by enforcing a positive coupling. This 
corresponds to including non-linear terms in the oscillator potential and/or 
the coupling, and hence modifies the model itself. This clearly makes sense 
from the point of view of modeling real materials. Including nonlinear terms
that strictly enforce a positive sign appears to be complicated with our 
simulation scheme, and therefore the standard method is likely preferable in 
cases where this would be necessary. Note, however, that including ``simple''
nonlinear terms in the oscillator potential should be possible, as long
as they do not cause a sign problem.

It is now a simple matter to include Monte Carlo updates involving the 
operators $\hat H_{a,b}$ with $a = 3,\ldots ,8$ in the 
Heisenberg simulation scheme developed in the previous Section. As before, 
we denote by $\hat H_{0,0}$ a
unit operator, not part of the Hamiltonian, introduced in order to fix the 
length of the index sequence to $L$. The operator string matrix element in
the weight function, Eq.~(\ref{wtau}), is more complicated than before, 
since the matrix elements of the phonon operators depend on the occupation 
numbers $n_b$;
\begin{eqnarray}
a_b|n_b\rangle & = &  \sqrt{n_b}|n_b-1\rangle, \nonumber \\
a^\dagger_b|n_b\rangle & =  & \sqrt{n_b+1}|n_b+1\rangle.
\end{eqnarray}
Hence, changes in the matrix element of $T_L$ have to be evaluated
in pair substitutions involving phonon operators. Apart from this, the methods
we use for updating the configurations are very similar to the Heisenberg 
case, and we shall therefore merely list the types of substitutions carried 
out and only very briefly comment on their implementation. 

As before, the update changing the expansion power $n$ is 
$[0,0] \leftrightarrow [1,b]$, with acceptance probabilities given
by Eq.~(\ref{accdia}) [with $J_0$ in place of $\Delta$]. The pair 
substitutions can be divided into two 
classes. A substitution of the first type is only accompanied by
changes in the spin states. We use the following ones:
\begin{mathletters}
\begin{eqnarray}
& & [1,b],[1,b] \leftrightarrow [2,b],[2,b] \label{spupd2} \\
& & [3,b],[4,b] \leftrightarrow [5,b],[6,b] \label{spupd3a} \\
& & [4,b],[3,b] \leftrightarrow [6,b],[5,b]. \label{spupd3b}
\end{eqnarray}
\end{mathletters}
These updates are again carried out with $T_L$ partitioned into one out of four
sets of subsequences. The other updates cause changes only in the phonon
states. These are
\begin{mathletters}
\begin{eqnarray}
& & [2,b],[2,b] \leftrightarrow [3,b],[4,b] \label{spupd4} \\
& & [3,b],[4,b] \leftrightarrow [4,b],[3,b] \label{spupd5} \\
& & [1,b],[3,b] \leftrightarrow [5,b],[2,b] \label{spupd6} \\
& & [1,b],[4,b] \leftrightarrow [6,b],[2,b] \label{spupd7} \\
& & [2,b],[5,b] \leftrightarrow [3,b],[1,b] \label{spupd8} \\
& & [2,b],[6,b] \leftrightarrow [4,b],[1,b] \label{spupd9} \\
& & [5,b],[6,b] \leftrightarrow [7,b],[8,b] \label{spupd10} 
\end{eqnarray}
\end{mathletters}
as well as the ones obtained from these by permuting operators both on the 
left and the right. Since the oscillators are independent (i.e., we have
not included phonon-phonon couplings), these updates can be carried out 
independently for all bonds, with $T_L$ partitioned into $N$ subsequences.

For small systems, we also carry out ``ring'' updates changing the winding 
number according to Eq.~(\ref{ringupd}). Global flips of single spins can 
again be carried out only at relatively high temperatures, thus necessitating
the use of a canonical ensemble for the spin sector at lower temperatures.
For the phonons, the algorithm is automatically grand canonical at any
temperature, since the phonon number is not conserved. We do, however, also 
include global fluctuations of the number of phonons at each bond. For this 
purpose, we monitor (during equilibration) the average number of phonons 
per oscillator, $\langle n_b \rangle$, and attempt adding or subtracting 
between $1$ and $\sqrt{\langle n_b \rangle}$ phonons at a time. 

Since the number of phonons is unbounded (in practice the largest number
sampled is limited, but can be very large at high temperatures), we cannot 
test the QMC program against exact diagonalization results with the full 
phonon space included. However, we have compared results with exact 
diagonalizations for a 4-site chain with the number of phonons per oscillator
restricted to less than or equal to two. This restricted case already 
encompasses all the elements of the simulation algorithm, and hence we can 
conclude that our program generates results exact within statistical errors.

\subsection{Results}

We shall here only present some selected illustrative results for the 
spin-Peierls model, deferring to a separate publication a detailed discussion 
of the physics of the model, and its relevance in interpreting various 
experimental results for quasi-1D spin systems.\cite{awsdkc} The main
purpose here is to give an impression about the kind of detailed information 
that can be extracted using the new QMC algorithm. We do, however, also point 
out some important model features of experimental relevance that can be 
inferred already from the results obtained here.

We note that in the adiabatic limit $\omega_0 \to 0$, the model should 
dimerize for any $\alpha$.\cite{cross} The situation for finite frequency is 
still not completely settled. For the case of noninteracting spinless 
fermions, which is equivalent to the $XY$ spin chain, Hirsch and Fradkin 
found numerical evidence that a critical phonon coupling is required for a
dimerization to occur. We here consider a phonon frequency $\omega_0=J/10$, 
and a coupling $\alpha = J/4$. Our low-temperature results show that 
the system is dimerized at $T=0$ for these parameters. Based on their exact 
diagonalization study including the $q=\pi$ phonons only, Augier and Poilblanc
\cite{augier} suggested a coupling $\alpha /J \approx 0.38$ (after adjusting
for a difference $\sqrt{2}$ in definitions) and a bare frequency $\omega_0/J 
= 0.3$ for describing $\alpha '-$NaV$_{\rm 2}$O$_{\rm 5}$. Hence, the case
considered here is closer to the adiabatic limit, and the spin-phonon coupling
is slightly weaker. This is still a regime that can be expected to be of 
relevance for real materials. As we will see, our bare coupling $J$ is in 
fact renormalized to a slightly higher value due to $q=0$ phonons, so
that our effective phonon frequency and coupling are somewhat lower in units 
of the effective spin-spin coupling.

We present simulation results for a wide range of finite temperatures, 
including low enough $T$ to give the ground state for all practical purposes. 
Systems with up to $128$ sites are considered, and a simulation typically 
consisted of $1-2 \times 10^7$ MC steps. At high temperatures 
($T/J \ge 0.125$),
we used the grand canonical ensemble for the spins, and at lower $T$ we
had to restrict ourselves to the canonical ensemble with $m^z=0$, due to
the low acceptance rate for global spin flips. For the relatively large
lattices considered, the effects of this restriction are minor. As already
noted, the simulation algorithm is automatically grand canonical for
the phonons, and we have found no practical problems with the ability of
the phonon numbers to fluctuate effectively.

The spin-phonon coupling causes an average uniform displacement 
$\langle x_i\rangle > 0$ of the oscillators, due to 
the energy lowering realized by increasing the 
average coupling, $J \to J + \alpha \langle x_i\rangle = J_{\rm eff}$, 
balanced by the increased potential energy of the oscillator. Hence, the 
spin-Peierls system is characterized by a temperature dependent effective 
spin-spin coupling $J_{\rm eff}(T)$. Fig.~\ref{figjeff} shows our QMC results 
for the effective coupling, along with its RMS fluctuation, defined as 
\begin{equation}
\sigma (J_{\rm eff}) = 
\alpha \sqrt{\langle x_i^2\rangle - \langle x_i\rangle ^2 } .
\end{equation}
As already noted, the relevance of the model to real materials is likely
limited to the regime where $J_{\rm eff}$ rarely fluctuates to negative 
values. The results of Fig.~\ref{figjeff} indicate that, for the model 
parameter considered here, this is the case for $T/J \alt 1$, where 
$J _{\rm eff} > \sigma (J_{\rm eff})$. 

At low temperatures, the lonq-wavelength spin susceptibility shows the 
behavior expected for a dimerized spin chain. Fig.~\ref{spxqfig} shows 
results for $\chi (q)$ for $q < \pi/2$. The behavior changes rapidly between 
inverse temperatures $\beta=16$ and $64$, from having only a very weak 
$q$-dependence in the $q \to 0$ regime in the former case, to a rapid 
suppression in the latter case. This clearly shows the presence of a spin gap.
There is only a minor change going to $\beta=128$, indicating that the system 
is essentially in its ground state at this inverse temperature. In order to 
accurately extract the uniform susceptibility $\chi (q \to 0)$, it is 
necessary to extrapolate to $q=0$ using $2-4$ low-$q$ points. This procedure 
is of sufficient accuracy in the interesting intermediate temperature regime 
where the susceptibility drops rapidly.

The uniform susceptibility is one of the most important experimentally
measurable quantities. It is often used to estimate the Heisenberg spin-spin 
coupling. In comparing results for the spin-Peierls model with the Heisenberg 
chain, it is natural to measure the temperature in units of the effective 
coupling at $T=0$, which we denote $J_{\rm eff}(0)$. We here have 
$J_{\rm eff}(0) \approx 1.273$ [see Fig.~\ref{figjeff}]. Susceptibilities 
calculated in the simulation, where the bare exchange $J=1$ is used to set the
temperature scale, then also
have to be adjusted by this factor since the definition (\ref{statsusc}) 
contains the inverse temperature. Hence, unless stated otherwise, we now 
define $\beta = J_{\rm eff}(0)/T$, and give the susceptibility in units of 
$1/J_{\rm eff}(0)$.

In Fig.~\ref{spsusc} we compare the uniform susceptibility of the 
spin-Peierls model with the exact Heisenberg result.\cite{eggert} There is 
a significant shift in both the peak position and the amplitude. Both can be 
understood on the basis of the reduced antiferromagnetism due to the 
fluctuations induced by the phonons. There are no signs of a suppression of 
the susceptibility for $T/J_{\rm eff}(0) \agt 0.1$, below which there is 
a sudden, very rapid drop. Above this drop, we find that the shape of the 
temperature dependence of $\chi$ can be quite well fit to the result for the 
Heisenberg chain without phonons, but with an exchange $J_{\rm fit} < 
J_{\rm eff}(0)$, as also indicated in the figure. The over-all magnitude of 
the susceptibility is then found to be lower than for a Heisenberg chain with
this exchange $J=J_{\rm fit}$. In terms of an effective Landee $g$-factor 
(i.e., the value of the $g$-factor one would deduce under the incorrect 
assumption that the system is described by a Heisenberg chain), which is 
$2$ for an ideal $S=1/2$ Heisenberg chain, this implies $g_{\rm fit} < 2$. 
For the parameters used here, we find $J_{\rm fit} \approx 0.82J$ and
$g_{\rm fit} \approx 1.86$. This is interesting, since experimentally it 
is sometimes found that the $g$-factor extracted from the susceptibility fit 
is less than $2$, whereas one would in fact expect \cite{gfactor} a $g$-factor
slightly larger than $2$. For example, Eggert was able to fit well 
experimental susceptibility data for Sr$_{\rm 2}$CuO$_{\rm 3}$ with the 
Heisenberg $\chi (T)$,\cite{eggert2} but the $g$-factor corresponding to the 
fit is as low as $1.6$, and this seeming discrepancy has not been explained. 
Based on our results presented here, we propose that the apparent $g$-factor 
reduction may be at least partially due to phonons. Additional calculations 
for various phonon frequencies and couplings are underway to further 
investigate this possibility. Note that the susceptibility of the 
spin-Peierls model deviates significantly from the Heisenberg $\chi (T)$ also 
if one considers a temperature-dependent energy scale $J_{\rm eff}(T)$ instead 
of the fixed scale $J_{\rm eff}(0)$ used above, as shown in the inset of 
Fig.~\ref{spsusc}. The shift in the peak position remains, and there is a 
regime where the phonons cause a significant susceptibility enhancement. This 
shows that the effects of the phonons cannot be simply captured by a
mapping to a Heisenberg chain with a temperature-dependent exchange 
equal to the average coupling $J_{\rm eff}(T)$ of the model, i.e., the 
fluctuations in $J_{\rm eff}(T)$ must be taken into account. These issues 
will be discussed in more detail elsewhere.\cite{awsdkc}

The effects of the phonons on
the antiferromagnetism can be seen directly in the staggered 
susceptibility, graphed in Fig.~\ref{spxpifig}. For the Heisenberg chain 
$\chi (\pi) = D{\rm ln}^{1/2}(\Lambda/T)/T$ at low temperatures, where the 
constants have been estimated as $D \approx 0.32$ and $\Lambda \approx 6$. 
\cite{sss} For the gapped spin-Peierls model we should have 
$\chi (\pi) \to {\rm constant}$. We find here that there is a maximum in 
$\chi (\pi)$ approximately at the same temperature at which the rapid drop in 
the uniform susceptibility is seen [the staggered structure factor $S(\pi)$
exhibits a similar maximum]. This is behavior is distinctively different from 
that of a gapped system with fixed values of the exchange couplings, such as a 
statically dimerized chain, for which $\chi (\pi)$ simply saturates at a 
value set by the gap. In the presence of dynamic phonons the staggered 
correlations among the instantaneous spin-spin couplings only build up 
gradually as $T$ is lowered, and therefore the antiferromagnetic spin 
correlations can initially  grow stronger than what they will eventually be 
in the statically dimerized $T=0$ state. Experimentally, this may have 
implications for NMR experiments, which probe the low-frequency dynamic
spin-spin correlations. The NMR rates for models including phonons should be 
accessible using the same numerical techniques (QMC and maximum-entropy 
analytic continuation) that have recently been used for the standard Heisenberg
chain.\cite{nmr,sss}

The phonon correlations are also interesting, and can give some indirect 
information on the dynamics as well. We define the static phonon structure 
factor and susceptibility according to
\begin{mathletters}
\begin{eqnarray}
S_x(q) & = & {\alpha^2\over N} \sum\limits_{j,l} {\rm e}^{-i(j-l)q}
\langle (x_{j}-\langle x\rangle)(x_{i}-\langle x\rangle) \rangle , \\
\chi_x(q) & = & {\alpha^2\over N} \sum\limits_{j,l} {\rm e}^{-i(j-l)q}
\int_0^\beta d\tau 
\langle (x_{j}(\tau)-\langle x\rangle)(x_{i}(0)-\langle x\rangle) \rangle .
\label{xxsx}
\end{eqnarray}
\end{mathletters}
We expect both to develop peaks at $q=\pi$ at low temperatures, signaling 
the dimerization instability. Fig.~\ref{xxfig} shows results at several 
temperatures. At high temperatures, both $S_x(q)$ and $\chi_x(q)$ are almost 
independent of $q$ and increase with increasing temperature. This reflects 
the behavior of (almost) independent harmonic oscillators. At lower
temperatures, the expected peaks at $q=\pi$ develop. In the 
high and intermediate temperature regime ($T \agt \omega_0$) graphed in 
Fig.~\ref{xxfig}, $ T\chi_x (q) =  S_x(q)$ within statistical errors. Hence, 
at these temperatures the phonon dynamics is for all wavelengths dominated 
by the first Matsubara frequency and is thus essentially classical. Note 
that the phonon susceptibilities have considerably
smaller statistical errors than the structure factors. This is contrary to
the situation for the spin structure factor and susceptibility, which are
calculated using the diagonal $S^z_i$ operators. This counterintuitive
feature of the QMC method can be traced to the simple form of the 
off-diagonal susceptibility estimator, Eq.~(\ref{offsusc}).

The static phonon structure factor and susceptibility can be related to the
dynamic real-frequency phonon correlation function (or spectral function)
$A(q,\omega)$ through sum rules. We define the spectral function as
\begin{equation}
A(q,\omega) = \alpha^2 \sum\limits_{j,l} {\rm e}^{iq(j-l)}
\int_{-\infty}^\infty dt {\rm e}^{-i\omega t}
\langle x_l(t)x_j (0) \rangle .
\end{equation}
From the Lehmann representation one can derive the following sum rules in
the standard way:\cite{hohenberg}
\begin{mathletters}
\begin{eqnarray}
S_x(q) & = & 
{1\over\pi} \int_0^\infty (1+{\rm e}^{-\beta\omega}) A(q,\omega) \\
\chi_x(q) & = & 
{2\over\pi} \int_0^\infty (1-{\rm e}^{-\beta\omega}) {1\over\omega}A(q,\omega).
\end{eqnarray}
\end{mathletters}
At $T=0$, these sum rules can be used to obtain an {\it upper bound for
the lowest phonon excitation of momentum $q$}, according to
\begin{equation}
w_{\rm min} (q) \le 2S_x(q)/\chi_x (q) .
\end{equation}
In the case of a single sharp phonon mode, $2S_x(q)/\chi_x (q)$ is the exact 
excitation energy. This bound has recently been used to extract spin and 
charge velocities of 1D electronic models from QMC data.\cite{cuo} Here
we demonstrate its use for the phonon spectrum. Fig.~\ref{figbound} shows
results for a 128-site system at a low temperature, $T=J/128$. For long
wavelengths, the calculated bound is at the bare phonon frequency 
$\omega_0/J=0.1$ within statistical errors, indicating a very weak effective 
coupling of these phonons to the spin system. For $q \to \pi$, there is a 
clear reduction, as expected due to the softening of the $q=\pi$ mode in a 
system that spontaneously dimerizes. In an infinite system, the $T \to 0$ 
bound $2S_x(\pi)/\chi_x (\pi) \to 0$ since a static dimerization implies
the classical relation $\chi_x (\pi) = \beta S_x(\pi)$. For a finite system, 
there is always a gap (decreasing with increasing size) to the lowest phonon 
excitation and both $S_x(q)$ and $\chi_x (q)$ therefore saturate at finite
values for all $q$. For $q \not= \pi$, the results shown in 
Fig.~\ref{figbound} are saturated, but $\chi (\pi)$ (but not $S_x(\pi)$) 
still grows with decreasing $T$ and the actual $T=0$ bound is therefore 
considerably lower than that in the figure.

We now calculate the size of the Peierls distortion. In the dimerized
state the average displacement alternates between even and odd sites;
$\langle x_i \rangle = \langle x \rangle \pm \delta$. Hence, the $T=0$
staggered phonon structure factor is, for large system sizes 
$N$, given by
\begin{equation}
S_x (\pi) = \alpha ^2 \delta^2 N .
\label{spidimer}
\end{equation}
Fig.~\ref{figdimer} shows $S_x (\pi)$ versus the inverse temperature
$J/T$ for several system sizes. As expected, 
the saturation takes place at a temperature 
which decreases with increasing system size. The results for $N=32$ and 
$N=128$ obey Eq.~(\ref{spidimer}) quite accurately, the ratio of the saturated 
values being $\approx 4$. For $N=8$ the value is considerably lower, 
indicating much larger fluctuations in this small system. Using the $N=128$ 
result with Eq.~(\ref{spidimer}) gives a dimerization $\alpha\delta \approx 
0.13$, i.e., the effective alternating average exchange is $J_i = J_{\rm eff}
(0)[1 \pm \Delta_J]$, with $\Delta_J \approx 0.10$. 

The main purpose here has been to illustrate how various physical quantities 
are accessible with the new QMC algorithm. We have therefore not discussed
how our results compare to, e.g., mean-field theory and previous numerical 
calculations including only the $q=\pi$ phonon mode.\cite{augier} These
important issues will be addressed in a future publication.

\section{Summary and Discussion}

In this paper
we have introduced a new quantum Monte Carlo algorithm based on the standard
perturbation expansion in the interaction representation. This starting point 
was first suggested by Prokof`ev {\it et al}.\cite{prokofev} Our 
implementation of the sampling of the series is different and is essentially
an adaptation of procedures previously developed for the Stochastic Series 
Expansion algorithm.\cite{sse2} We have shown that the SSE sum and the 
continuous imaginary time path integral are in fact very strongly related 
to each other. 

As in the SSE scheme, the central element of the new method is the ordered
sequence of operators. In the interaction representation formulation 
developed here, time enters in the form of auxilliary variables associated 
with the sequence positions. The operator updates leading to changes in the 
particle propagation paths (kink-antikink creation and annihilation, in the 
language of Prokof`ev {\it et al.}\cite{prokofev}) are carried out via 
substitutions of pairs of off-diagonal and constant operators, with the time 
field held fixed. An efficient procedure for collective updating of whole 
segments of the time field was introduced. We also derived expressions for 
several types of important operator expectation values, and compared these 
with the corresponding expressions previously obtained within the SSE scheme.

Not having carried out comparisons with the procedures suggested by Prokof`ev 
{\it et al.}, we do not know which of the two sampling schemes is more 
efficient --- most likely this is model dependent. We believe that the
method discussed here is more efficient than the SSE approach in cases where
the diagonal part of the Hamiltonian dominates the internal energy. However,
the weight function is simpler in the SSE case, and again it is not easy
to make a general statement of exactly when the advantages of new method 
become significant. 

We note that the loop-cluster algorithm, invented by Evertz {\it et al.},
\cite{evertz} has recently been used with considerable success in studies of 
various spin-1/2 Heisenberg systems,\cite{loopstudies} in particular in the 
exact formulation developed by Beard and Wiese.\cite{beard} Variants of the 
loop algorithm have also been suggested for the 1D Hubbard model,\cite{naoki1}
and for $S > 1/2$ spin chains.\cite{naoki2} However, physics results for these
cases have not yet been presented. Although its performance for the Heisenberg
model can be considered spectacular (however, for some quantities more accurate
results have actually been obtained with the SSE method\cite{sseparam}), it 
is not clear how the loop algorithm will fare with more complicated models,
such as spin-phonon models. It is known to break down completely in some 
cases.\cite{evertz} In contrast, the stochastic series expansion algorithm 
and the perturbation series schemes are in principle completely general and 
are practically useful for a wide range of models for which the sign problem 
can be avoided. 

As a demonstration of the power of the new method, we have implemented it 
for studies of a spin-Peierls model. We considered oscillators associated
with the bonds, coupled to the spins via a linear modulation of the exchange. 
Unlike earlier studies of 1D electronic models coupled to phonons,
\cite{hirsch} we used the occupation number basis also for the phonons. 
The new QMC algorithm is ideally suited for this type of models, since the 
bare phonon part of the Hamiltonian is diagonal. The initial results presented
here for the spin-Peierls chain indicate that reliable results can be
obtained with modest computer resources (the simulations were run on 
mid-range workstations). The method should be very useful for resolving 
issues related to the effects of dynamic phonons on the physics of the 
recently discovered inorganic spin-Peierls compounds, as well as other 
quantum spin systems. Work along these lines is in progress.\cite{awsdkc} 
1D itinerant electronic models of the Hubbard and $t$-$J$ types including 
phonons can also be studied using the procedures developed here. We also 
believe that studies of spin-phonon models in higher dimensions are 
feasible with our new method.

The results presented here for the spin-Peierls model already indicate some 
important consequences of dynamic phonons at finite temperature. The type
of phonons considered here naturally lead to a temperature dependent 
effective spin-spin coupling. We 
found that the uniform magnetic susceptibility still has a shape rather 
similar to that of the Heisenberg chain in a sizable regime close to the 
susceptibility maximum often used to extract the size of the exchange 
coupling from experimental data. However, both the value of $J$ and the 
$g$-factor extracted from a fit are reduced relative to a Heisenberg chain with
a coupling equal to the average coupling of the spin-phonon model. We propose 
that this dynamic effect may at least partially be the reason for the reduced 
$g$-factor found in some quasi-1D systems.\cite{eggert2} Furthermore, these 
results cast some doubts on the validity of detailed extractions 
\cite{spdiag1,spdiag3} 
of the nearest-neighbor and next-nearest-neighbor couplings in GeCuO$_{\rm 3}$ 
from fits of exact diagonalization data of frustrated Heisenberg chains to 
susceptibility measurements. It is likely that the couplings extracted from 
such fits do not directly correspond to the true spin-spin couplings of the 
system, but are influenced by the temperature dependence of the couplings as 
well as their fluctuations, as discussed above. Interchain couplings likely
also have some non-negligible effects on the susceptibility. Unfortunately, 
the QMC approach introduced here does not allow for studies of frustrated 
systems (at least not at very low temperatures), due to sign problems. 
Since $\alpha '$-NaV$_{\rm 2}$O$_{\rm 5}$ is not expected to be frustrated,
\cite{nadiscovery,naexperiments} we believe that detailed experimental studies 
of this material in combination with finite-$T$ QMC studies will be of key 
importance in clarifying the microscopic physics of the spin-Peierls 
materials.

\section{Acknowledgments}

This work is supported by the National Science Foundation under Grants 
DMR-89-20538 (AWS and DKC) and DMR-96-16574 (RRPS).

\begin{table}
\begin{tabular}{ldddd}
~~ $q/\pi$   & ~~~~~~~~ $S(q)$ (QMC)    & ~~~~~~$S(q)$ (exact) 
& ~~~~~~~~ $\chi(q)$ (QMC)    & ~~~~~~$\chi(q)$ (exact) 

\\ 
\hline
~~   0   &  0.008581(24)   & 0.008590  &  0.06864(20)  &   0.068720   \\
~~  1/6  &  0.048110(5)    & 0.048120  &  0.11822(3)   &   0.118266   \\
~~  1/3  &  0.102865(8)    & 0.102875  &  0.14422(3)   &   0.144236   \\
~~  1/2  &  0.169906(13)   & 0.169917  &  0.19746(5)   &   0.197478   \\
~~  2/3  &  0.263758(18)   & 0.263768  &  0.32669(8)   &   0.326648   \\
~~  5/6  &  0.43371(5)     & 0.433742  &  0.8027(3)    &   0.803118   \\
~~    1  &  0.95472(18)    & 0.954566  &  3.9897(13)   &   3.989397   \\
\end{tabular}
\vskip2mm
\caption{Simulation results for the static structure factor and the static 
susceptibility of a 12-site Heisenberg chain at $\beta=8$, compared with
the exact results. The numbers within parentheses indicate the statistical
errors (defined as one standard deviation of the averages), i.e. $0.123(45)$ 
stands for $0.123 \pm 0.045$.}
\label{tab1}
\end{table}

\begin{figure}
\caption{A configuration generated for an 8-site periodic system with 
$\Delta=1$ at $\beta=4$. A row represents a spin state 
$|\alpha (p)\rangle$, with $p=0$ 
to $p=L$ from top to bottom. Solid and open circles indicate up and 
down spins, respectively. Dashed and solid bars represent operators $[1,b]$ 
and  $[2,b]$, and their associated times are graphed to the right.}
\label{conf}
\end{figure}

\begin{figure}
\caption{Upper panel: The uniform magnetic susceptibility (in units
of $1/J$) of a 128-site Heisenberg chain calculated using the new QMC 
method (circles with error bars) compared with the exact result for the 
infinite system (curve). Lower panel: The relative deviation 
$(\chi_{\rm QMC} - \chi_{\rm exact})/\chi_{\rm exact}$ 
of the QMC data from the exact result. The solid and open circles
are for grand canonical and canonical simulations, respectively.} 
\label{suscfig}
\end{figure}

\begin{figure}
\caption{Upper panel: The truncation $L$ vs.~the number of MC steps performed 
during the initial parts of two simulations of a 128 site system at $\beta=8$.
In one case (solid curve), all the times were simultaneously updated,
whereas in the other case (dashed curve) the times were consecutively
updated one-by-one.}
\label{equifig}
\end{figure}

\begin{figure}
\caption{The autocorrelation function for $E$ (solid squares), $\chi (0)$
(open squares), $\chi (\pi)$ (solid circles), and $S(\pi)$ (open circles),
obtained in a simulation of a 128-site system at inverse temperature 
$\beta = 8$.}
\label{autofig}
\end{figure}

\begin{figure}
\caption{The temperature dependence of the effective spin-spin coupling 
(upper panel), and the size of its RMS fluctuation (lower panel).} 
\label{figjeff}
\end{figure}

\begin{figure}
\caption{The spin susceptibility (in units of $1/J$) vs.~wave-number in the 
long-wavelength regime for a system with $N=128$ at inverse temperatures 
$\beta=16$ (solid circles), $\beta=32$ (open circles), $\beta=64$ (solid 
squares), and $\beta=128$ (open squares). The statistical errors are at most 
the size of the symbols.}
\label{spxqfig}
\end{figure}

\begin{figure}
\caption{The uniform susceptibility [in units of $1/J_{\rm eff}(0)$] 
vs.~temperature [in units of $J_{\rm eff}(0)$], compared with the exact 
Heisenberg result (solid curve). For the Heisenberg chain $J_{\rm eff}=J=1$. 
The dashed curve is the Heisenberg susceptibility for 
$J=0.86 \times J_{\rm eff}(0)$ 
and a $g$-factor $\approx 1.82$. The inset shows the QMC data graphed on the 
temperature scale set by the temperature dependent coupling $J_{\rm eff}(T)$ 
[$\chi$ here also contains this factor, i.e., it is given in units of 
$1/J_{\rm eff}(T)$], compared with $\chi (T)$ for the Heisenberg chain.}
\label{spsusc}
\end{figure}

\begin{figure}
\caption{The staggered susceptibility [in units of $1/J_{\rm eff}(0)$]
vs.~temperature (solid circles), compared with results for the Heisenberg 
chain (open circles). The curve indicates the asymptotic divergent form 
for the Heisenberg chain. }
\label{spxpifig}
\end{figure}

\begin{figure}
\caption{The static phonon structure factor vs.~wave number at $T/J=1.0$ (open
circles), $T/J=0.5$ (solid circles), $T/J=0.25$ (open squares), and
$T/J=0.125$ (solid squares). The corresponding susceptibilities, multiplied
by $T/J$, are indicated by the (in some cases barely visible) solid curves. 
The statistical errors of the susceptibilities are considerably smaller 
than those of the structure factors, and are not indicated for sake of 
clarity. The dashed lines indicate the $q$-independent structure factors 
of independent harmonic oscillators at the corresponding temperatures.}
\label{xxfig}
\end{figure}

\begin{figure}
\caption{Upper bound (in units of the bare exchange $J$) for the lowest 
phonon excitation vs.~momentum for a 128-site system at $T=J/64$. The 
dashed line indicates the bare, momentum independent phonon frequency 
$\omega_0/J=0.1.$}
\label{figbound}
\end{figure}

\begin{figure}
\caption{The staggered phonon structure factor vs.~inverse temperature for 
systems of size $N=8$ (solid circles), $N=32$ (open circles), and $N=128$ 
(solid squares). The statistical errors are smaller than the symbols.}
\label{figdimer}
\end{figure}

\end{document}